\documentclass[useAMS,usenatbib,referee]{biom}
\def\bSig\mathbf{\Sigma}

\usepackage[figuresright]{rotating}
\usepackage{mathtools}
\usepackage{graphicx}
\usepackage{epstopdf}
\usepackage{xfrac}
\usepackage{adjustbox}
\usepackage{booktabs}
\usepackage{amsmath}
\newcommand\numberthis{\addtocounter{equation}{1}\tag{\theequation}}

\usepackage{float}
\usepackage{caption}
\usepackage{setspace}

\title[]{Sample Size Estimation using a Latent Variable Model for Mixed Outcome Co-Primary, Multiple Primary and Composite Endpoints}

\author{Martina McMenamin\\%$^{*}$\email{martina.mcmenamin@mrc-bsu.cam.ac.uk} \\
	  MRC Biostatistics Unit, University of Cambridge, Cambridge, UK \and 
	   Jessica K. Barrett \\
	  MRC Biostatistics Unit, University of Cambridge, Cambridge, UK
	  \and
	  Anna Berglind \\
	  Late RIA, R\&D BioPharmaceuticals, AstraZeneca, Gothenburg, Sweden
	  \and 
	  James M.S. Wason \\
	    MRC Biostatistics Unit, University of Cambridge, Cambridge, UK\\
	    Population Health Sciences Institute, Newcastle University, Newcastle, UK}

\begin{document}

%\date{{\it Received September} 2019. {\it Revised September} 2019.\newline 
%{\it Accepted September} 2019.}

%\pagerange{\pageref{firstpage}--\pageref{lastpage}} \pubyear{2019}

%\volume{59}
%\artmonth{September}
%\doi{10.1111/j.1541-0420.2005.00454.x}

%  This label and the label ``lastpage'' are used by the \pagerange
%  command above to give the page range for the article

\label{firstpage}

%  pub the summary here

\begin{abstract}
Mixed outcome endpoints that combine multiple continuous and discrete components to form co-primary, multiple primary or composite endpoints are often employed as primary outcome measures in clinical trials. There are many advantages to joint modelling the individual outcomes using a latent variable framework, however in order to make use of the model in practice we require techniques for sample size estimation. In this paper we show how the latent variable model can be applied to the three types of joint endpoints and propose appropriate hypotheses, power and sample size estimation methods for each. We illustrate the techniques using a numerical example based on the four dimensional endpoint in the MUSE trial and find that the sample size required for the co-primary endpoint is larger than that required for the individual endpoint with the smallest effect size. Conversely, the sample size required for the multiple primary endpoint is reduced from that required for the individual outcome with the largest effect size. We show that the analytical technique agrees with the empirical power from simulation studies. We further illustrate the reduction in required sample size that may be achieved in trials of mixed outcome composite endpoints through a simulation study and find that the sample size primarily depends on the components driving response and the correlation structure and much less so on the treatment effect structure in the individual endpoints.
\end{abstract}

%
%  Please place your key words in alphabetical order, separated
%  by semicolons, with the first letter of the first word capitalized,
%  and a period at the end of the list.
%


\maketitle

\section{Introduction}
Sample size estimation plays an integral role in the design of a study. The objective is to determine the minimum sample size that is large enough to detect, with a specified power, a clinically meaningful treatment effect. Although it is crucial that investigators have enough patients enrolled to detect this effect, overestimating the sample size also has ethical and practical implications. Namely, in a placebo-controlled trial, more patients are subjected to a placebo arm than is necessary, therefore withholding access to potentially beneficial drugs from them and delaying access to future patients. Furthermore it results in longer, more expensive trials, using resources that could be allocated elsewhere. \\
One vital aspect of sample size determination is the primary endpoint. Usually this is a single outcome, however in some instances there may be multiple outcomes of equal relevance. Various combinations of multiple outcomes may be selected as the primary endpoint, depending on the hypothesis of interest. One option is a co-primary endpoint, which requires that a treatment effect must occur in all of the selected primary outcomes in order for the treatment to be classed as effective overall. Alternatively, multiple primary endpoints may be of interest and thus overall effectiveness may be concluded if a treatment effect occurs in at least one of the outcomes. Another possibility is a composite endpoint, involving some combination of the outcomes which reduces to a one dimensional outcome used for inference. In composite responder endpoints, the outcomes are combined by labelling patients as `responders' or `non-responders' based on whether they exceeded predefined thresholds in each of the outcomes, collapsing the information contained in the components into an overall binary, ordinal or categorical response indicator. \\
For each of these endpoints, the individual outcomes may be a mix of multiple continuous, ordinal and binary measures. In this case, there is a lack of an obvious multivariate distribution for the joint outcomes. One possible way to jointly model the outcomes is using a latent variable framework arising in the graphical modelling literature, which assumes that the discrete outcomes are latent continuous variables subject to estimable thresholds, and proceed with a multivariate normal distribution \citep{multprobit, multprobit2}. By employing this framework we can take account of the correlation between the outcomes, improve the handling of missing data in individual components and potentially increase efficiency. Furthermore, in the case of multiple primary outcomes, it may reduce the severity of multiple testing corrections required. \\
A key barrier of benefiting from these advantages is the lack of techniques for sample size determination. If a mixed outcome composite is selected as the primary endpoint in a trial then the sample size calculation is typically based on the overall binary responder endpoint, analysed using a standard binary method such as logistic regression. Sample size calculations performed in this way are valid but when applying the latent variable model in this context it is desirable to avail of the greatly increased power, hence requiring a technique based on the novel joint modelling approach \citep{McMenamin2019}. In the case of co-primary and multiple primary endpoints, the trials are typically powered for the continuous component, see for example \citep{PREMIER}. The co-primary and multiple primary outcomes proposed may also be restricted to being on the same scale. In this case standard multivariate distributions can be used for which some methods have been proposed, however these methods are not routinely implemented. \\
A recent and comprehensive overview of the existing literature for sample size determination in clinical trials with multiple endpoints is provided by \cite{samplesizebook}. The review found many proposals for power and sample size calculations for multiple continuous outcomes. Some of these suggestions were based on assuming that the endpoints were bivariate normally distributed (\cite{Sozu, Xiong}) and extended for the case of more than two endpoints (\cite{Sozu2, Sugimoto}). Other work focused on testing procedures and found two-sample t-tests, which reject only if each t-statistic is significant, to be conservative and biased (\cite{Eaton, Julious}). Other efforts were focused on investigating and controlling the type I error rate (\cite{Senn, Chuang, Kordzakhia, Hung}). All of these methods focus on the requirement of effects on all endpoints. Methods for effects on at least one endpoint also exist (\cite{Senn, Hung, Dmitrienko, Gong}).\\
Less comprehensive consideration has been given to the case of multiple binary endpoints. Five methods of power and sample size calculation based on three association measures are introduced for co-primary binary endpoints by \cite{Sozu3}. Sample size calculation for trials using multiple risk ratios and odds ratios for treatment effect estimation is discussed by \cite{Hamasaki}. \cite{Song} explores co-primary endpoints in non-inferiority clinical trials. Consideration has also been given to the case where two co-primary endpoints are both time-to-event measures where effects are required in both endpoints (\cite{Hamasaki2, Sugimoto3, Sugimoto4}) and at least one of the endpoints (\cite{Sugimoto2}).\\
Even less consideration has been given to the mixed outcome setting. One paper considers overall power functions and sample size determinations for multiple co-primary endpoints that consist of mixed continuous and binary variables (\cite{Sozu4}). They assume that response variables follow a multivariate normal distribution, where binary variables are observed in a dichotomized normal distribution, and use Pearson's correlations for association. A modification was suggested to this method by \cite{Wu} which involved using latent-level tests and pairwise correlations and provided increased power. These methods focus on the co-primary endpoint case, where effects are required in all outcomes. The case of multiple primary or composite responder endpoints where the components are measured on different scales has not been considered, each of which will require distinct hypotheses. \\
In this paper we will build on the work by \cite{Sozu4} and \cite{Wu} for co-primary continuous and binary endpoints to include any combination of continuous, ordinal and binary outcomes for co-primary, multiple primary and composite endpoints. The paper will proceed as follows. We will introduce the latent variable model and detail how it can be used in each context. We will illustrate the appropriate hypothesis tests for each of the three combinations of mixed outcomes and propose power calculation and sample size estimation techniques. We illustrate the methods on a four dimensional endpoint consisting of two continuous, one ordinal and one binary outcome using a numerical example based on the MUSE trial \citep{MUSE}. We further investigate empirically the behaviour of the power and sample size when employing this approach for a composite outcome and compare it with the standard binary method for a range of scenarios and data structures. Finally we conclude with a discussion and recommendations for practice.  

\section{Latent Variable Model}

We begin by constructing the model in \cite{McMenamin2019}. Let $n_{T}$ and $n_{C}$ represent the number of patients in the treatment group and the control group respectively and let K be the number of outcomes measured for each patient. Let $\mathbf{Y_{Ti}}=(Y_{Ti1},...,Y_{TiK})^{T}, i=1,...,n_{T}$ be vector of K responses for patient $i$ on the treatment arm and $\mathbf{Y_{Ci}}=(Y_{Ci1},...,Y_{CiK})^{T}, i=1,...,n_{C}$ the vector of K responses for patient i on the control arm. Without loss of generality, the first $1\leq k \leq k_{m}$ elements of $\mathbf{Y_{Ti}}$ and $\mathbf{Y_{Ci}}$ are observed as continuous variables, the next $k_{m}< k \leq k_{o}$ are observed as ordinal and the remaining $k_{o}< k \leq K$ are observed as binary. We use the biserial model of association by \cite{Tate}, which is based on latent continuous measures manifesting as discrete variables. Formally, we say that $\mathbf{Y_{Ti}}$ and $\mathbf{Y_{Ci}}$ have latent variables $\mathbf{Y_{Ti}^{*}}$ and $\mathbf{Y_{Ci}^{*}}$ respectively, where $\mathbf{Y_{Ti}^{*}}\sim N_{K}(\boldsymbol{\mu_{T}},\Sigma_{T})$ and $\mathbf{Y_{Ci}^{*}}\sim N_{K}(\boldsymbol{\mu_{C}},\Sigma_{C})$

\begin{align*} 
\Sigma_{T}=\begin{pmatrix}
\sigma_{T1}^{2} & \ldots & \rho_{T1K}\sigma_{T1}\sigma_{TK} \\
\vdots & \ddots & \vdots \\
\rho_{T1K}\sigma_{T1}\sigma_{TK} & \cdots & \sigma_{TK}^{2} \\
\end{pmatrix}, &&
\Sigma_{C}=\begin{pmatrix}
\sigma_{C1}^{2} & \ldots & \rho_{C1K}\sigma_{C1}\sigma_{CK} \\
\vdots & \ddots & \vdots \\
\rho_{C1K}\sigma_{C1}\sigma_{CK} & \cdots & \sigma_{CK}^{2} \\
\end{pmatrix}
\end{align*}\\
For $k\neq k': 1 \leq k < k' \leq K$ let $Var(Y_{Tik})=\sigma_{Tk}^{2}, Var(Y_{Cik})=\sigma_{Ck}^{2}, Corr(Y_{Tik},Y_{Tik'})=\rho_{Tkk'}, Corr(Y_{Cik},Y_{Cik'})=\rho_{Ckk'}$, where $\rho_{Tkk'}$ and $\rho_{Ckk'}$ are the association measures between the endpoints. Then, for:

\begin{itemize}
\item $1\leq k \leq k_{m}: Y_{Tik}=Y_{Tik}^{*}$ and $Y_{Cik}=Y_{Cik}^{*}$
\item $k_{m}< k \leq k_{o}: $ \newline

$Y_{Tik}=\begin{cases}
0 & \text{if }  \tau_{k0}\leq Y_{Tik}^{*}<\tau_{k1}, \\
  1& \text{if } \tau_{k1}\leq Y_{Tik}^{*}<\tau_{k2},\\
\vdots & \hspace{12mm} \vdots \\
w_{k} & \text{if } \tau_{kw_{k}}\leq Y_{Tik}^{*}<\tau_{k(w_{k}+1)}
\end{cases} 
Y_{Cik} = \begin{cases}
0 & \text{if }  \tau_{k0}\leq Y_{Cik}^{*}<\tau_{k1}, \\
  1& \text{if } \tau_{k1}\leq Y_{Cik}^{*}<\tau_{k2},\\
\vdots & \hspace{12mm} \vdots \\
w_{k} & \text{if } \tau_{kw_{k}}\leq Y_{Cik}^{*}<\tau_{k(w_{k}+1)}
\end{cases}$
\vspace{3mm}

\item $k_{o} < k \leq K: Y_{Tik} = \begin{cases}
  0 & \text{if } \tau_{k0}\leq  Y_{Tik}^{*}< \tau_{k1}, \\
  1 & \text{if } \tau_{k1}\leq Y_{Tik}^{*}< \tau_{k2}
\end{cases}
\hspace{5mm} Y_{Cik} = \begin{cases}
  0 & \text{if } \tau_{k0}\leq  Y_{Cik}^{*}< \tau_{k1}, \\
  1 & \text{if } \tau_{k1}\leq Y_{Cik}^{*}< \tau_{k2}
\end{cases}$
\end{itemize}

\vspace{5mm}

We set $\tau_{k0}=-\infty, \tau_{k(w_{k}+1)}=\infty $ and the intercepts equal to zero for $k_{m}< k \leq k_{o}$ in order to estimate the cut-points. Additionally, $\tau_{k0}=-\infty, \tau_{k1}=0, \tau_{k2}=\infty $ for $k_{o}< k \leq K$ so that the intercepts can be estimated for the outcomes observed as binary. Furthermore, for $k_{m} < k \leq K$, $\sigma_{Tk}^{2}=\sigma_{Ck}^{2}=1$. We assume that $\Sigma=\Sigma_{T}=\Sigma_{C}$ and partition $\Sigma= \left(\begin{array}{cc}
\Sigma_{11} & \Sigma_{12}\\
\Sigma_{21} & \Sigma_{22}
\end{array}\right)$ so that $\Sigma$ is as shown in (\ref{Sigmafull}).

\begin{equation}\label{Sigmafull}
\Sigma = \left(\begin{array}{cccccc}
\sigma_{1}^{2} & \cdots & \rho_{1k_{m}}\sigma_{1}\sigma_{k_{m}}&\rho_{1k_{m+1}}\sigma_{1} & \cdots &  \rho_{1K}\sigma_{1} \\
\vdots & \ddots &\vdots & \vdots &\ddots & \vdots\\
\rho_{1k_{m}}\sigma_{1}\sigma_{k_{m}} & \cdots & \sigma_{k_{m}}^{2} & \rho_{k_{m}k_{m+1}}\sigma_{k_{m}}& \cdots & \rho_{k_{m}K}\sigma_{k_{m}}\\
\rho_{1k_{m+1}}\sigma_{1} & \ldots & \rho_{k_{m}k_{m+1}}\sigma_{k_{m}} & 1 &  \ldots &  \rho_{k_{m+1}K}\\
\vdots & \ddots &\vdots & \vdots &\ddots & \vdots\\
\rho_{1K}\sigma_{1} & \cdots & \rho_{k_{m}K}\sigma_{k_{m}} & \rho_{k_{m+1}K} & \cdots & 1\\
\end{array}\right)
\end{equation}

For $k_{m} < k \leq K$ we can define the conditional mean for outcome $k$ as $\mu_{k|1...k_{m}}=\mu_{k}^{*}+\Sigma_{12}\Sigma_{22}^{-1}(y_{k}-\mu_{k})$ where $\mu_{k}^{*}$ is the latent mean. The correlation matrix for the outcomes can then be defined using the pairwise correlations between elements of $Y_{i}$ and $Y_{i}^{*}$ as below. 

\begin{equation}
\Gamma=\begin{pmatrix}
\boldsymbol{D}^{-\frac{1}{2}}\Sigma_{11}\boldsymbol{D}^{-\frac{1}{2}} & \boldsymbol{D}^{-\frac{1}{2}}\Sigma_{12}\\
& \Sigma_{22}
\end{pmatrix}
\end{equation}

where $\boldsymbol{D}^{-\frac{1}{2}}=diag(\sigma_{1}^{-1},...,\sigma_{k_{m}}^{-1})$.\\
This framework can be used as the basis for which to jointly model multiple continuous, ordinal and binary outcomes. However, the endpoints of interest may differ, with distinct hypotheses for each, thus requiring different techniques for estimating the sample size. We present approaches for sample size determination in the case of mixed co-primary endpoints, multiple primary endpoints and composite endpoints below, highlighting the different assumptions and requirements in each case.  

\section{Sample Size Estimation}

\subsection{Co-primary Endpoints}
One potential application of the latent variable model is to mixed outcome co-primary endpoints. In this case, a treatment must be shown to be effective as measured by each of the outcomes in order to be deemed effective overall. \cite{Sozu4} propose a method to calculate the sample size for a mixture of continuous and binary endpoints using this framework, which was amended by \cite{Wu} to use latent level tests for increased power. We generalise this work to include ordinal outcomes, as shown below. 

\subsubsection{Hypothesis Testing}

In many clinical trials the hypothesis of interest is based on superiority, namely that the proposed treatment will perform better than the control treatment, specified by some predefined margin. The null hypothesis is that the difference in treatment effects for the treatment arm and control arm is zero. This is straightforward to formalise in the case of one endpoint but less so when there are multiple co-primary endpoints, particularly when they are measured on different scales. The hypothesis of interest is as shown in (\ref{coprimhypth}). 
\begin{equation}\label{coprimhypth}
\begin{aligned}
H_{0}: & \hspace{3mm}\exists \hspace{0.3mm} k \hspace{1mm} s.t.\hspace{1mm} \pi_{Tk}-\pi_{Ck} \leq 0\\
H_{1}: & \hspace{3mm} \pi_{Tk}-\pi_{Ck}>0 \hspace{1mm} \forall k
\end{aligned}
\end{equation}

For $k_{o}<k\leq K$ we can specify $\pi_{Tik}=P(Y_{Tik}=0)=P(Y_{Tik}^{*}<0)$ and $\pi_{Cik}=P(Y_{Cik}=0)=P(Y_{Cik}^{*}<0)$ for the treatment and control group respectively. We can generalise this assumption to account for the ordinal endpoints based on the fact that for $k_{m}<k\leq k_{o}$ $\pi_{Tik}=P(Y_{Tik}=w_{k})=P(\tau_{kw_{k}}<Y_{Tik}^{*}<\tau_{k(w_{k}+1)})$. Therefore, multiple levels in the ordinal outcomes can be considered by selecting the appropriate $\tau$ thresholds. For instance, $\pi_{Tik}=P(Y_{Tik}=0)+P(Y_{Tik}=1)+P(Y_{Tik}=2)=P(-\infty<Y_{Tik}^{*}<\tau_{k3})$. As the latent means are estimable by maximum likelihood, $\mu_{Ti1}=\Phi^{-1}(\pi_{Ti1}),\ldots,\mu_{TiK}^{*}=\Phi^{-1}(\pi_{TiK})$ in the treatment group and $\mu_{Ci1}=\Phi^{-1}(\pi_{Ci1}),\ldots,\mu_{CiK}^{*}=\Phi^{-1}(\pi_{CiK})$ in the control group. \\
We can proceed by specifying that the hypothesis in (\ref{coprimhypth}) holds if and only if the hypothesis
\begin{equation}\label{newhyp}
\begin{aligned}
H_{0}^{*}: & \hspace{3mm} \exists  \hspace{0.3mm} k \hspace{1mm} s.t. \hspace{1mm} \delta_{k}^{*} \leq 0 \hspace{1mm} \\ H_{1}^{*}:  & \hspace{3mm} \delta_{k}^{*}>0 \hspace{1mm} \forall k
\end{aligned}
\end{equation}

\noindent holds, where $\delta_{k}^{*}=\mu_{Tk}^{*}-\mu_{Ck}^{*}$, as highlighted by \cite{Wu}. The maximum likelihood estimates $\hat{\mu}_{Tk}^{*}$ and $\hat{\mu}_{Ck}^{*}$ can be used for a test of $H_{0}^{*}$ and the variance of this test statistic can be obtained using the inverse of the Fisher information matrix. 

\subsubsection{Overall Power}
Having specified the hypothesis to include ordinal outcomes, the power in this case is as defined for mixed continuous and binary co-primary endpoints \citep{Wu}, as shown in (\ref{coprimpower}). 
\begin{equation}\label{coprimpower}
1-\beta=P\left(\bigcap_{k=1}^{k_{m}}\lbrace Z_{k} > z_{\alpha}\rbrace\bigcap_{k_{m+1}}^{K} \lbrace Z^{*}_{k} > z_{\alpha}\rbrace \mid \boldsymbol{\delta}\right) \simeq P\left(\bigcap_{k=1}^{K}\lbrace Z_{k}^{\dagger}>z_{k}^{\dagger}\rbrace \mid \boldsymbol{\delta} \right)
\end{equation}

for $\boldsymbol{\delta}=(\delta_{1},...,\delta_{k_{m}},...,\delta_{k_{o}},...,\delta_{K})^{T}\neq\mathbf{0}$ and 

\begin{equation}
Z_{k}^{\dagger}=\begin{cases} Z_{k}-\dfrac{\delta_{k}}{\sigma_{k}}\sqrt{\dfrac{\kappa n_{T}}{1+\kappa}}=\dfrac{\bar{Y}_{Tk}-\bar{Y}_{Ck}-\delta_{k}}{\sigma_{k}\sqrt{\dfrac{1+\kappa}{\kappa n_{T}}}}, 	 k=1,...,k_{m}\\
\\
Z_{k}^{*}-\delta_{k}^{*}\sqrt{\dfrac{\kappa n_{T}}{1+\kappa}}=\dfrac{\hat{\mu}_{Tk}^{*}-\hat{\mu}_{Ck}^{*}-\delta_{k}^{*}}{\sqrt{\dfrac{1+\kappa}{\kappa n_{T}}}}, 	 k=k_{m+1},...,K 
\end{cases}
\end{equation}

\vspace{5mm}

\begin{equation}
z_{k}^{\dagger}=\begin{cases} z_{\alpha}-\dfrac{\delta_{k}}{\sigma_{k}}\sqrt{\dfrac{\kappa n_{T}}{1+\kappa}}, k=1,...,k_{m}\\
\\
 z_{\alpha}-\delta_{k}^{*}\sqrt{\dfrac{\kappa n_{T}}{1+\kappa}}, k=k_{m+1},...,K
 \end{cases}
\end{equation}

\noindent where $\delta_{k}=\mu_{Tk}-\mu_{Ck}$, $\delta_{k}^{*}=\mu_{Tk}^{*}-\mu_{Ck}^{*}$, $\kappa=n_{C}/n_{T}$ and $z_{\alpha}$ is the $(1-\alpha)100^{th}$ standard normal percentile. Therefore, the power can be approximated as shown in (\ref{powercalc}).

\begin{equation}
1-\beta \simeq P\left(\bigcap_{k=1}^{K}\lbrace Z_{k}^{\dagger}>z_{\alpha}^{\dagger}\rbrace \mid \boldsymbol{\delta} \right) = \Phi_{K}\left(-z_{1}^{\dagger},...,-z_{K}^{\dagger};\Gamma\right)\label{powercalc}
\end{equation}

\subsubsection{Sample Size Calculation}
Assuming $n_{T}=n_{C}=n$ it is possible, as discussed by \cite{samplesizebook}, to rearrange (\ref{powercalc}) to obtain a sample size formula in terms of n as shown in (\ref{sampn}),
%\begin{equation}
%1-\beta \leq \int_{z_{1-\alpha}}^{\infty}\ldots\int_{z_{1-\alpha}}^{\infty}f\left(z_{1},\ldots,z_{k_{m}},z_{k_{m+1}}^{*},\ldots,z_{K}^{*};\sqrt{nk}\boldsymbol{\delta^{\dagger}},\Gamma\right)dz_{1},\ldots,dz_{K}^{*}
%\end{equation}
%where $\boldsymbol{\delta^{\dagger}}=\left(\dfrac{\delta_{1}}{\sigma_{1}},\ldots,\dfrac{\delta_{K}}{\sigma_{K}}\right)$. 
\begin{equation}
n=\dfrac{\left(C_{k}+z_{1-\alpha}\right)^{2}}{k\delta_{k}^{2}}\label{sampn}
\end{equation}
\noindent where $C_{k}$ is the solution of
\begin{equation}
1-\beta=\int_{-\infty}^{\gamma_{1}C_{k}+z_{1-\alpha}(\gamma_{1}-1)}\ldots\int_{-\infty}^{\gamma_{k-1}C_{k}+z_{1-\alpha}(\gamma_{k-1}-1)}\int_{-\infty}^{C_{k}}f\left(z_{1},\ldots,z_{K}^{*};\mathbf{0},\Gamma\right)dz_{K}^{*}\ldots dz_{1}
\end{equation}

\subsection{Multiple Primary Endpoints}
Multiple primary endpoints are distinct from co-primary endpoints in that the treatment needs to be shown to work in at least one of the outcomes in order to be classed as effective overall. We would expect the sample size required to be reduced compared with the co-primary endpoint case which would require power to detect treatments in all outcomes. We can extend the work of \cite{Sozu4} and \cite{Wu} to allow for sample size estimation for multiple primary endpoints. 

\subsubsection{Hypothesis Testing}
The hypothesis of interest, accounting for the fact that a significant effect in only one outcome is required, is shown below. 
\begin{equation}\label{multprimhypth}
\begin{aligned}
H_{0}: & \hspace{3mm} \pi_{Tk}-\pi_{Ck} \leq 0 \hspace{1mm} \forall k\\
H_{1}: & \hspace{3mm} \exists \hspace{0.3mm} k \hspace{1mm} s.t.\hspace{1mm} \pi_{Tk}-\pi_{Ck} > 0
\end{aligned}
\end{equation}

\noindent As before, $\pi_{Tk}$ and $\pi_{Ck}$ can be determined for $k_{m}<k \leq k_{o}$ using the relevant $\tau$ thresholds. 
\begin{equation}\label{newmulthyp}
\begin{aligned}
H_{0}^{*}: & \hspace{3mm}  \hspace{3mm} \delta_{k}^{*} \leq 0 \hspace{1mm} \forall k \\ 
H_{1}^{*}:  & \hspace{3mm} \exists  \hspace{0.3mm} k \hspace{1mm} s.t. \hspace{1mm} \delta_{k}^{*} > 0
\end{aligned}
\end{equation}

\noindent The difference in latent means $\delta_{k}^{*}=\mu_{Tk}^{*}-\mu_{Ck}^{*}$ and their variance are estimated using the maximum likelihood estimates and Fisher information matrix respectively, as before. 

\subsubsection{Overall Power}
The overall power can then be specified as shown below
\begin{equation}
1-\beta=P\left(\bigcup_{k=1}^{k_{m}}\lbrace Z_{k} > z_{\alpha}\rbrace\bigcup_{k_{m+1}}^{K} \lbrace Z^{*}_{k} > z_{\alpha}\rbrace \mid \boldsymbol{\delta}\right) \simeq P\left(\bigcup_{k=1}^{K}\lbrace Z_{k}^{\dagger}>z_{k}^{\dagger}\rbrace \mid \boldsymbol{\delta} \right)\\
\end{equation}

\noindent where $Z_{k}^{\dagger}$ and $z_{k}^{\dagger}$ is as defined for co-primary endpoints and $n_{T}=n_{C}=n$. In order to obtain an appropriate power function we rely on the inclusion-exclusion principle. 
\begin{align*}
 P\left(\bigcup_{k=1}^{K}\lbrace Z_{k}^{\dagger}>z_{k}^{\dagger}\rbrace \mid \boldsymbol{\delta} \right)=\sum_{k=1}^{K}P\left(\{Z_{k}^{\dagger}>z_{k}^{\dagger}\} \mid \boldsymbol{\delta} \right)-\sum_{k<l}P\left(\{Z_{k}^{\dagger}>z_{k}^{\dagger}\} \cap \{Z_{l}^{\dagger}>z_{l}^{\dagger}\} \mid \boldsymbol{\delta} \right) \\ \hspace{30mm} +\sum_{k<l<m}P\left(\{Z_{k}^{\dagger}>z_{k}^{\dagger}\} \cap \{Z_{l}^{\dagger}>z_{l}^{\dagger}\}  \cap \{Z_{m}^{\dagger}>z_{m}^{\dagger}\} \mid \boldsymbol{\delta} \right) \\ \hspace{-40mm} + \ldots + (-1)^{K-1} \sum_{k< \ldots <K} P\left(\bigcap_{k=1}^{K}\{Z_{k}^{\dagger}>z_{k}^{\dagger}\}\mid \boldsymbol{\delta}\right)
\end{align*}

A closed form expression for the overall power is shown in (\ref{powermultprim}). 
\begin{equation}
 P\left(\bigcup_{k=1}^{K}\lbrace Z_{k}^{\dagger}>z_{k}^{\dagger}\rbrace \mid \boldsymbol{\delta} \right)=\sum_{i=1}^{K}\left((-1)^{i-1} \sum_{I\subseteq\{1,\ldots,K\}}P\left(\bigcap_{k \in I} \{Z_{k}^{\dagger}>z_{k}^{\dagger}\} \mid \boldsymbol{\delta}\right)\right)\label{powermultprim}
\end{equation}

Assuming $n_{T}=n_{C}=n$, we can input different values for n to achieve the required power.

\subsection{Composite Endpoints}
\label{compendsection}
A review conducted by \cite{Wason2019} showed that composite responder endpoints are widely used. They identified many clinical areas in which they are common, such as oncology, rheumatology, cardiovascular and circulation. As shown by \citet{McMenamin2019}, the latent variable framework may be used to model the underlying structure of these mixed outcome composite endpoints to greatly improve efficiency. The joint distribution of the continuous, ordinal and binary outcomes is modelled using the latent variable structure as before. However, in this case the endpoint of interest is a composite responder endpoint and so the required quantity is some function of the probability of response in the treatment group $p_{T}$ and in the control group  $p_{C}$. \\
For instance, an overall responder index $S_{i}$ can be formed for patient $i$, where $S_{i}=1$ if $Y_{i1} \leq \eta_{1}, \ldots, Y_{iK}^{*} \leq \eta_{K}$ and 0 otherwise, where the quantities $(\eta_{1},\ldots,\eta_{K})$ are predefined responder thresholds. Generalisations where response just requires a certain number of the components to meet the thresholds are possible, but involve more complex sums. We can specify $p_{iT}$ and $p_{iC}$, the probability of response for patient $i$ in the treatment arm and control arm respectively, as shown in (\ref{probcomp}),
\begin{equation}
\begin{aligned}
p_{iT}=P(S_{i}=1|T_{i}=1)= & \int_{-\infty}^{\eta_{1}}\ldots\int_{-\infty}^{\eta_{K}}f_{Y_{1},\ldots,Y_{K}}(y_{i1},\ldots,y_{iK}|T_{i}=1,\boldsymbol{\theta})dy_{K}\ldots dy_{1} \\
p_{iC}=P(S_{i}=1|T_{i}=0)= & \int_{-\infty}^{\eta_{1}}\ldots\int_{-\infty}^{\eta_{K}}f_{Y_{1},\ldots,Y_{K}}(y_{i1},\ldots,y_{iK}|T_{i}=0,\boldsymbol{\theta})dy_{K}\ldots dy_{1}
\end{aligned}\label{probcomp}
\end{equation}

\noindent  where $\boldsymbol{\theta}$ is the vector of model parameters and we assume that $p_{T} \sim N(\delta_{T}, \sigma_{\delta_{T}})$ and $p_{C} \sim N(\delta_{C}, \sigma_{\delta_{C}})$. As in the case of co-primary and multiple primary endpoints, the assumptions allow us to estimate latent means $(\mu_{k_{m+1}}^{*},\ldots,\mu_{K}^{*})$ for the observed discrete components using the model parameters. 

\subsubsection{Hypothesis Testing}
In the mixed outcome composite endpoint setting, note that although we are exploiting the latent multivariate Gaussian structure for efficiency gains we are ultimately still interested in a one dimensional endpoint, such as the difference in response probabilities between the treatment and control arms of the trial. This is distinct from the co-primary and multiple primary endpoints cases, where the overall hypothesis test must be based on some union or intersection of the hypotheses for the individual outcomes. For the composite endpoint we can formulate the hypothesis as shown in (\ref{hyp}),
\begin{equation}\label{hyp}
\begin{aligned}
H_{0}: & \hspace{3mm} p_{T}-p_{C} \leq 0\\
H_{1}: &  \hspace{3mm} p_{T}-p_{C} > 0
\end{aligned}
\end{equation}

\noindent where $p_{T}$ and $p_{C}$ are as in (\ref{probcomp}). For sample size estimation, we require the distribution of $\delta$ under $H_{1}$, which we can assume to be $\delta \sim N(\delta_{T}-\delta_{C},\sigma_{\delta}^{2})$. The hypothesis can therefore be stated as
\begin{equation}\label{hyp2}
\begin{aligned}
H_{0}: & \hspace{3mm} \delta^{*} \leq 0\\
H_{1}: &  \hspace{3mm} \delta^{*} > 0
\end{aligned}
\end{equation}

\noindent where $\delta^{*}=\delta_{T}^{*}-\delta_{C}^{*}$, $\delta_{T}^{*}= \Phi_{K}(\eta_{1},\cdots ,\eta_{K};\boldsymbol{\mu}_{T}^{*},\Sigma_{T})$, $\delta_{C}^{*} = \Phi_{K}(\eta_{1},\cdots ,\eta_{K};\boldsymbol{\mu}_{C}^{*},\Sigma_{C})$ and $\Phi_{K}(.;\boldsymbol{\mu},\Sigma)$ is the K-dimensional multivariate normal distribution function, with mean vector $\boldsymbol{\mu}$ and covariance matrix $\Sigma$. Estimates of the quantities can be obtained using the maximum likelihood estimates for the model parameters, as in the co-primary and multiple primary endpoint settings, so that $\hat{\delta}_{T}^{*}=\Phi_{K}(\eta_{1},\cdots ,\eta_{K};\boldsymbol{\hat{\mu}}_{T}^{*},\widehat{\Sigma}_{T})$ and $\hat{\delta}_{C}^{*}=\Phi_{K}(\eta_{1},\cdots ,\eta_{K};\boldsymbol{\hat{\mu}}_{C}^{*},\widehat{\Sigma}_{C})$, where $\boldsymbol{\mu_{T}}^{*}$ is the K-dimensional vector of mean values in the treatment arm and $\boldsymbol{\mu_{C}}^{*}$ is the corresponding vector for the control arm. Using a Taylor series expansion, we can obtain the quantity $\sigma_{\delta}^{2}$ as follows.
\begin{equation}\label{deltavar}
var(\hat{\delta}^{*}) \approx (\boldsymbol{''\delta})^{T}Cov(\boldsymbol{\hat{\theta}})(\boldsymbol{''\delta})
\end{equation}

\noindent Then, $\widehat{var}(\hat{\delta}^{*})=(\boldsymbol{''\delta_{T}})^{T}\widehat{Cov}(\boldsymbol{\hat{\theta}})(\boldsymbol{''\delta_{T}})$, where $\boldsymbol{''\delta}$ is the vector of partial derivatives of $\delta^{*}$ with respect to each of the parameter estimates. One potential difficulty for conducting sample size estimation using the latent variable model in practice is that the vector of model parameters $\boldsymbol{\theta}$ may be large depending on the number of outcomes. We can obtain $\boldsymbol{\hat{\theta}}$ and covariance matrix $\widehat{Cov}(\boldsymbol{\hat{\theta}})$ by fitting the model to pilot trial data. If all model parameters and their covariance matrix could be specified fitting the model on pilot data would not be required, however this would be difficult in practice, especially for more than two components. 

\subsubsection{Overall Power}
To test the hypothesis in (\ref{hyp}), we need to determine the critical value, cv. As the endpoint of interest is specified in terms of the overall one dimensional composite endpoint, we can use the formula assumed when employing the standard test of proportions technique and the approximation for the distribution of the test statistic $\delta$ under $H_{1}$ so that $cv = \hspace{1mm} \sigma_{\delta}z_{\alpha}$. As $\sigma_{\delta}=\sqrt{\tfrac{\sigma_{\delta_{T}}}{n_{T}}+\tfrac{\sigma_{\delta_{C}}}{n_{C}}}$, we can assume that $\sigma_{T}=\sigma_{C}=\sigma$ and $n_{T}=n_{C}=n$, so that $\delta \sim N(\delta_{T}-\delta_{C}, 2\sigma^{2}/n)$. The power is deduced in the standard way, as demonstrated below.
\begin{align*}
1-\beta = & \hspace{1mm} P \left( \bar{p}_{T}-\bar{p}_{C}  > z_{\alpha}\sqrt{\sfrac{2\sigma^{2}}{n}} \mid H_{1} \right) \\
= & \hspace{1mm} P\left(Z > \frac{z_{\alpha}\sqrt{\sfrac{2\sigma^{2}}{n}}-\delta^{*}}{\sqrt{\sfrac{2\sigma^{2}}{n}}} \mid H_{1} \right)\\
= & \hspace{1mm} \Phi\left(\frac{\delta^{*}}{\sqrt{\sfrac{2\sigma^{2}}{n}}}-z_{\alpha}\right)  \numberthis \label{comppower}
\end{align*}

\subsubsection{Sample Size Calculation}
Note that $\sigma_{\delta}^{2}$ estimated using (\ref{deltavar}) gives $\tfrac{2\sigma^{2}}{n}$, however to obtain a formula in terms of the required sample size we will need to separate $n$ from the variance estimate. By fitting the model to pilot trial data we can obtain an estimate for $\sigma^{2}$, as the value of $n$ will be known in this instance and $n$ can be obtained using (\ref{compsampsize}).
\begin{equation}
n = \hspace{1mm} \frac{2\sigma^{2}(z_{1-\beta}+z_{\alpha})^{2}}{\delta^{*2}}\label{compsampsize}
\end{equation}

\noindent This is similar to the sample size equation used for the binary method, however $\sigma$ will be different and $\delta^{*}$ is obtained using latent means as opposed to provided directly. 

\section{Numerical Application}
\subsection{MUSE trial}
We illustrate the technique for sample size determination using the MUSE trial \citep{MUSE}. The trial was a phase IIb, randomised, double-blind, placebo-controlled study investigating the efficacy and safety of anifrolumab in adults with moderate to severe systemic lupus erythematosus (SLE). Patients (n=305) were randomised (1:1:1) to receive anifrolumab (300mg or 1000mg) or placebo, in addition to standard therapy every 4 weeks for 48 weeks. The primary endpoint in the study was the percentage of patients achieving an SLE Responder Index (SRI) response at week 24, with sustained reduction of oral corticosteroids ($<$10mg/day and less than or equal to the dose at week 1 from week 12 through 24). SRI is comprised of a continuous Physician's Global Assessment (PGA) measure, a continuous SLE Disease Activity Index (SLEDAI) measure and an ordinal British Isles Lupus Assesment Group (BILAG) measure \citep{SRI4}. The study had a target sample size of 100 patients per group based on providing 88\% power at the 0.10 alpha level, to detect at least 20\% absolute improvement in SRI(4) response rate at week 24 for anifrolumab relative to placebo. The investigators assumed a 40\% placebo response rate. \\

\subsection{Model}
In this case $\left(Y_{1},Y_{2},Y_{3},Y_{4}\right)$ are the two continuous, ordinal and binary components respectively and $\left(Y_{1},Y_{2},Y_{3}^{*},Y_{4}^{*}\right) \sim N_{4}\left(\boldsymbol{\mu^{*}},\Sigma\right)$ where,
\begin{align}
\boldsymbol{\mu^{*}}=\left( \mu_{1},\mu_{2},\mu_{3}^{*},\mu_{4}^{*}\right)^{T} &&
\Sigma=\begin{pmatrix}
\sigma_{1}^{2} & \rho_{12}\sigma_{1}\sigma_{2} & \rho_{13}\sigma_{1} & \rho_{14}\sigma_{1} \\
  \rho_{12}\sigma_{1}\sigma_{2} & \sigma_{2}^{2} & \rho_{23}\sigma_{2} & \rho_{24}\sigma_{2} \\
\rho_{13}\sigma_{1} & \rho_{23}\sigma_{2} & 1 & \rho_{34}\\
\rho_{14}\sigma_{1} & \rho_{24}\sigma_{2}  & \rho_{34} & 1\\
\end{pmatrix}
\end{align} 
\noindent  and the ordinal and binary components may be related to their latent variables as shown in (\ref{MUSEdiscrete}). The thresholds $(\tau_{31},\tau_{32}, \tau_{33}, \tau_{34})$ are estimated from the data.
\begin{equation}
Y_{i3}=\begin{cases}
0 & \text{if }  -\infty < Y_{i3}^{*}<\tau_{31}, \\
  1& \text{if } \hspace{5mm} \tau_{31}\leq Y_{i3}^{*}<\tau_{32},\\
  2& \text{if }  \hspace{5mm} \tau_{32}\leq Y_{i3}^{*}<\tau_{33},\\
   3& \text{if }  \hspace{5mm} \tau_{33}\leq Y_{i3}^{*}<\tau_{34},\\
      4& \text{if }  \hspace{5mm} \tau_{34}\leq Y_{i3}^{*}<\infty,\\
\end{cases} \hspace{10mm}
Y_{i4}=\begin{cases}
  0 & \text{if } -\infty <  Y_{i4}^{*}< 0, \\
  1 & \text{if }  \hspace{8mm} 0 \leq Y_{i4}^{*}< \infty
\end{cases}\label{MUSEdiscrete}
\end{equation}

\noindent We can use the MUSE trial to design future studies where we assume that the endpoints of interest are co-primary, multiple primary and composite endpoints. The overall power functions for each are shown below.
\begin{align*}
& Power_{co}= \Phi_{4}\left(-z_{1}^{\dagger},-z_{2}^{\dagger},-z_{3}^{\dagger},-z_{4}^{\dagger};\Sigma\right)\\
& Power_{mult}= \sum_{i=1}^{4}\left((-1)^{i-1} \sum_{I\subseteq\{1,2,3,4\}}\Phi_{k \in I}\left( -z_{k}^{\dagger};\Sigma\right)\right)\\
& Power_{comp}= \Phi\left(-z\right) 
\end{align*}

\noindent where $z_{k}^{\dagger}=z_{\alpha}-\tfrac{\delta_{k}}{\sqrt{\sfrac{2\sigma_{k}^{2}}{n}}}$ for $k=\{1,2\}$ and $z_{k}^{\dagger}=z_{\alpha}-\tfrac{\delta_{k}^{*}}{\sqrt{\sfrac{2}{n}}}$ for $k=\{3,4\}$. In the composite setting $z=\frac{\delta^{*}}{\sqrt{\sfrac{2\sigma^{2}}{n}}}-z_{\alpha}$ where $\sigma$ is estimated using (\ref{deltavar}).

\newpage
\subsection{Results}
\begin{figure}[h]
\centering
\includegraphics[scale=0.9]{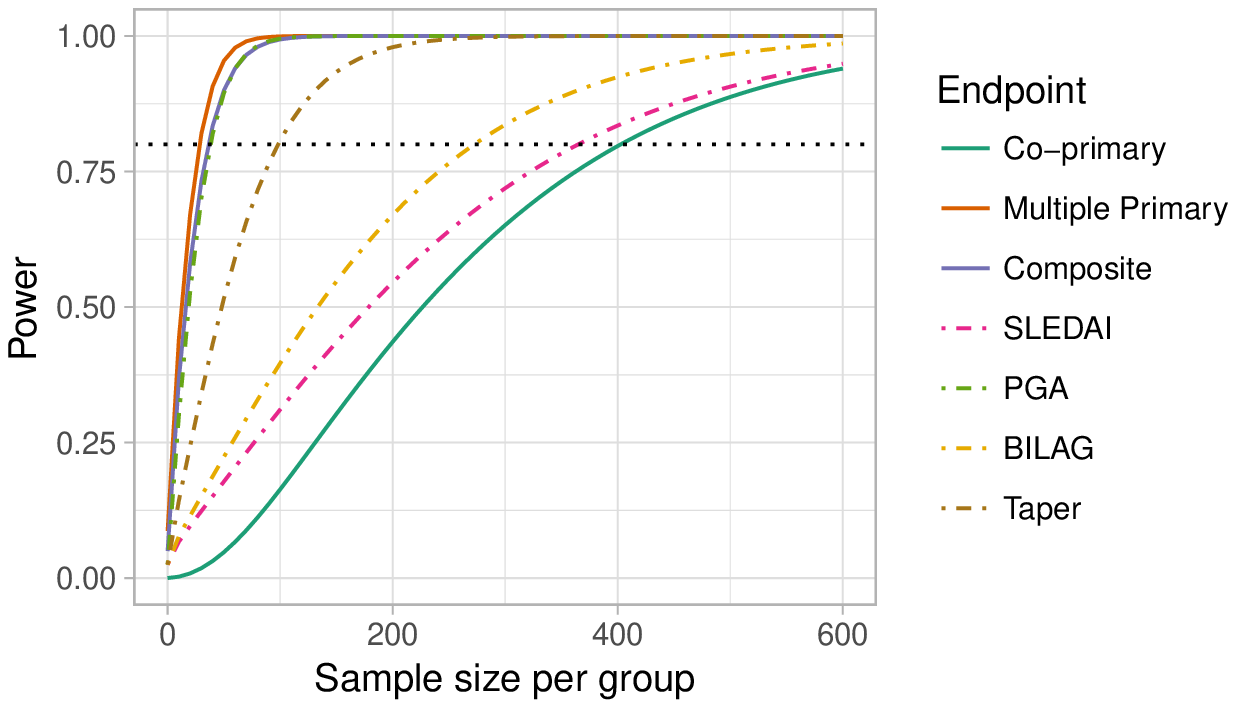}\caption[Power function for outcomes treated as co-primary, multiple primary and composite endpoints]{Power function for individual SLEDAI (continuous), PGA (continuous), BILAG (ordinal) and Taper (binary) outcomes and the power functions with when they are treated as co-primary, multiple primary and composite endpoints using data from the MUSE trial}\label{CoMultPrim}
\end{figure}

The power functions for all three joint endpoints are shown, along with that of the individual endpoints, in Figure \ref{CoMultPrim}. The multiple primary endpoints have the highest power, where $80\%$ is achieved for $n=29$. This is followed closely by the composite endpoint, which requires $n=37$. Note that the power function for the composite endpoint is almost identical to that of PGA, the component with the highest effect size. As we would expect the power is considerably lower for co-primary endpoints, which would require $n=400$ for $80\%$ power. 

\noindent Table \ref{MUSEcoprim} shows the sample sizes required in each group, for the co-primary and multiple primary endpoints to obtain an overall power of at least 80\% to detect a difference of 0.88 in SLEDAI, 0.38 in PGA, 0.24 in BILAG and 0.40 in the taper outcome at alpha level $\alpha=0.025$, based on the values observed in the trial. We also allow for uncertainty in the variance of the continuous measures by setting $\sigma_{1}^{2}=18,19,20$ and $\sigma_{2}^{2}=0.35,0.45,0.55,0.65$. The sample sizes required for each individual endpoint are also shown based on achieving a power of at least 80\%. Allowing for uncertainty in the variance of the SLEDAI outcome varies the required sample size for the co-primary endpoint but not the multiple primary endpoint. The opposite is true when the assumed variance of the PGA outcome is changed, namely affecting the sample size required for the multiple primary endpoint but not the co-primary. This is intuitive given that the treatment effect observed in the SLEDAI outcome is smallest and is largest for the PGA outcome. The effect on the power of removing an endpoint and altering the correlation between endpoints is shown in the Appendix.

\begin{table}\caption[Sample sizes for co-primary endpoints in the MUSE trial data]{Sample sizes $n=n_{C}=n_{T}$ for the co-primary and multiple primary endpoints for overall power $1-\beta \approx 0.80$, $\alpha=0.025$, $k_{2}=2, k_{o}=K=1$ using the MUSE trial data. $SS_{1},SS_{2},SS_{3},SS_{4}$ are sample sizes required per group for the individual endpoints for a power of at least $1-\beta=0.80$}
\label{MUSEcoprim}
\centering
\begin{adjustbox}{width=1\textwidth}
\begin{tabular}{cccccccccccccc}
\toprule
\multicolumn{2}{c}{SLEDAI} & \multicolumn{2}{c}{PGA} & \multicolumn{2}{c}{BILAG} & \multicolumn{2}{c}{Taper} & $n_{co}$ & $n_{mult}$ & $SS_{1}$ & $SS_{2}$ & $SS_{3}$ & $SS_{4}$ \\[0.5ex]
\cmidrule(r){1-2} \cmidrule(r){3-4} \cmidrule(r){5-6} \cmidrule(r){7-8} $\delta_{1}$ & $\sigma_{1}^{2}$  & $\delta_{2}$ & $\sigma_{2}^{2}$ & $(\pi_{T3},\pi_{C3})$ & $\delta_{3}^{*}$ & $(\pi_{T4},\pi_{C4})$ & $\delta_{4}^{*}$ &   & & & & &\\  \hline 
0.88 & 18 & 0.38 & 0.35 & (0.97,0.95) & 0.24 & (0.54,0.38)  & 0.40 & 403 & 29 & 365 & 39 & 273 & 99 \\ 
0.88 & 19 & 0.38 & 0.35 &  (0.97,0.95) & 0.24 & (0.54,0.38) & 0.40 & 419 & 29 & 386 & 39 & 273 & 99\\ 
0.88 & 20 & 0.38 & 0.35 &  (0.97,0.95)  & 0.24 & (0.54,0.38) & 0.40 & 435 & 29 & 406 & 39 & 273 & 99 \\ 
& & & & & & & & & & & \\
0.88 & 18 & 0.38 & 0.45 &  (0.97,0.95) & 0.24 & (0.54,0.38) & 0.40 & 403 & 34 & 365 & 49 & 273 & 99 \\ 
0.88 & 18 & 0.38 & 0.55 &  (0.97,0.95) & 0.24 & (0.54,0.38) & 0.40 & 403 & 39 & 365 & 60 & 273 & 99 \\ 
0.88 & 18 & 0.38 & 0.65 &  (0.97,0.95) & 0.24 & (0.54,0.38) & 0.40 & 403 & 42 & 365 & 71 & 273  & 99 \\ 
\hline
\end{tabular}
\end{adjustbox}
\end{table}

\noindent Table \ref{MUSEsamplesize} shows the sample size required per group assuming that a future trial in SLE is to be conducted using the composite responder endpoint, allowing for uncertainty in $\sigma$. The estimated variance for the risk difference from the trial dataset is $\sigma_{\delta}=0.048$ with correlation parameters $\rho_{12}=0.448, \rho_{13}=0.521, \rho_{14}=0.003, \rho_{23}=0.448, \rho_{24}=-0.031,\rho_{34}=0.066$. For a risk difference of 0.20, the required sample size per group is 20, compared to 100 for 88\% power in the standard binary method. Accounting for uncertainty in the latent variable treatment effect variance, $\sigma_{\delta}=0.10$ would increase the required sample size per group to 40, which is a more conservative estimate for use in practice. If the method were to be employed for increased power, rather than a decrease in required sample size, the estimated power of the latent variable method is over 99.99\% for sample sizes giving 88\% power at the 0.10 alpha level in the binary method. The empirical power is shown for the latent variable method in 1000 simulated datasets, which is approximately 88\% for each sample size, as required. 

\begin{table}[H]\caption[Sample sizes from the latent variable method for the MUSE trial data]{Sample sizes required using the latent variable method ($n_{lat}$) and the binary method ($n_{bin}$) for overall power $1-\beta \approx 88\%$, $\alpha=0.10$, $k_{2}=2, k_{o}=K=1$ to detect a response risk difference of 0.20, 0.18 and 0.16 as in the original MUSE trial. Estimated power is shown from the latent variable method for the sample size required by the binary method}
\label{MUSEsamplesize}
\centering
\begin{tabular}{cccccc}
\toprule
Risk difference & $\sigma$  & $n_{lat}$ & Empirical power (\%) & $n_{bin}$ & Power (\%)\\ [0.5ex]
 \midrule
0.20 & 0.05 & 20 & 88.05 & 100 & 99.99   \\
0.20 & 0.06 & 24 & 87.01& 100 & 99.99 \\
0.20 & 0.07 & 28 & 87.62 & 100 & 99.98  \\
0.20 & 0.08 & 32 & 87.04 & 100 & 99.96  \\
0.20 & 0.09 & 36 & 87.83 & 100 & 99.89  \\
0.20 & 0.10 & 40 & 88.12 & 100 & 99.89  \\
\bottomrule
\end{tabular}
\end{table}

\vspace{8mm}

\section{Behaviour of Power Function and Sample Size for Composite Endpoints}
We can further explore the properties of the functions proposed for composite endpoints empirically and consider how factors such as the correlation structure and drivers of response impact on the resulting power or sample size. As the primary purpose of using the latent variable model in this context is for efficiency gain, we will compare the power and sample size required using the technique we proposed with that required when the standard binary analysis is conducted, using only the response indicator for each patient \citep{McMenamin2019}. 

\subsection{One Continuous, One Ordinal, One Binary}

\begin{figure}
\centering
\includegraphics[scale=0.9]{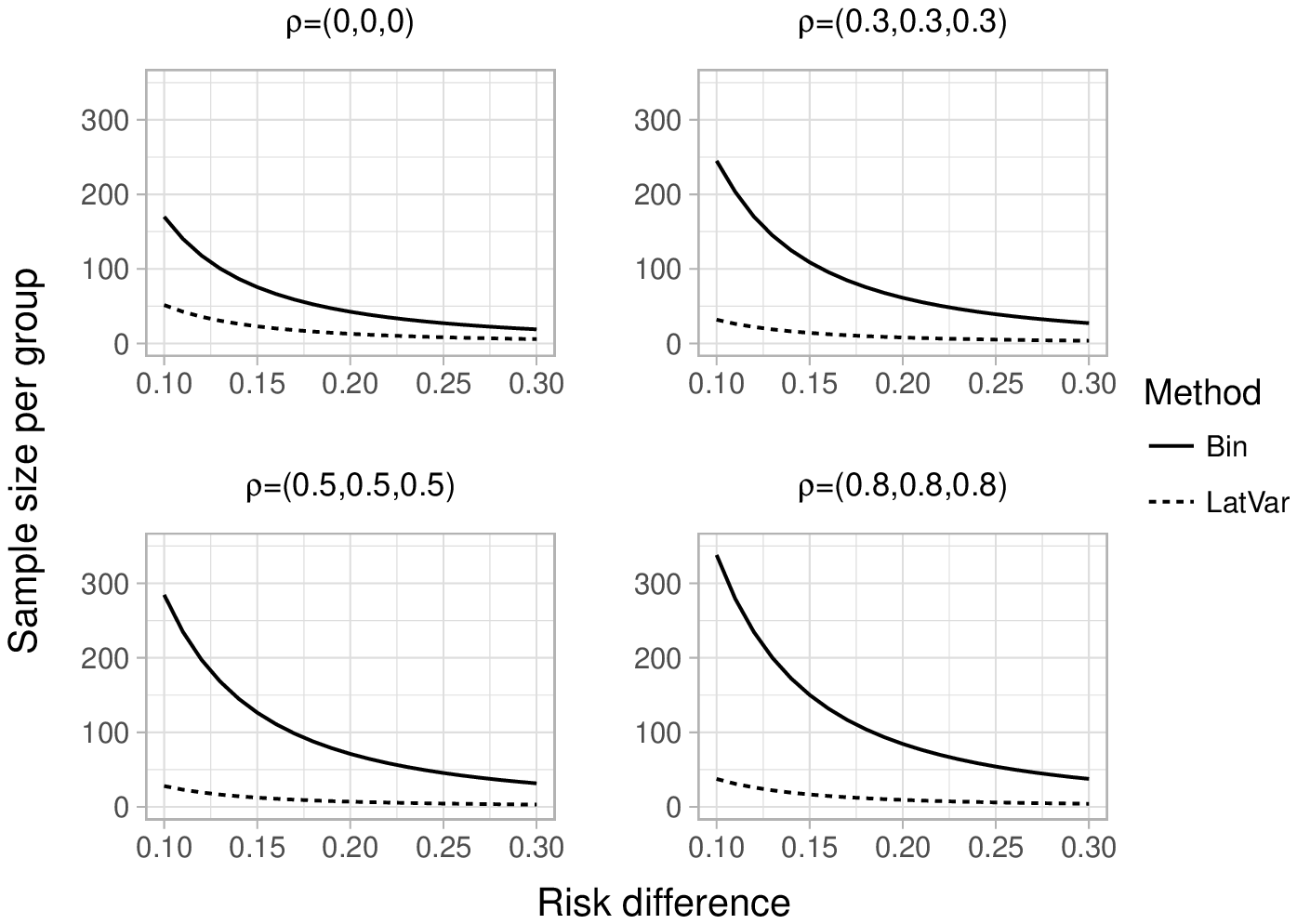}\caption[Sample size per group as risk difference changes using the latent variable and standard binary methods]{\setstretch{1.0}Estimated sample size per group for different values of the risk difference using the latent variable and standard binary methods when the composite endpoint is formed from one continuous, one ordinal and one binary outcome, where all components drive response and correlations between the outcomes are between 0 and 0.8, where $\rho=(\rho_{12},\rho_{13},\rho_{23})$}\label{changerisk}
\end{figure}

\noindent The probability of response in each arm, the overall power and sample size per arm is determined as illustrated in Section \ref{compendsection}. We begin by considering the case where the composite is a combination of one continuous, one ordinal and one binary outcome and the components are dichotomised at their mean so that all three are responsible for discriminating between responders and non-responders, i.e. all components drive response. We generate 1000 simulated datasets and fit the model to each, using the median variance estimate for $\sigma^{2}$. Figure \ref{changerisk} shows the change in estimated sample size per group as the overall treatment effect on the composite endpoint changes and everything else remains constant. This is shown for a range of correlations between the three endpoints. The sample size required for the latent variable method generally decreases slightly as the correlation between the endpoints increases, except when the correlation is already high. 

\begin{figure}
\centering
\includegraphics[scale=0.9]{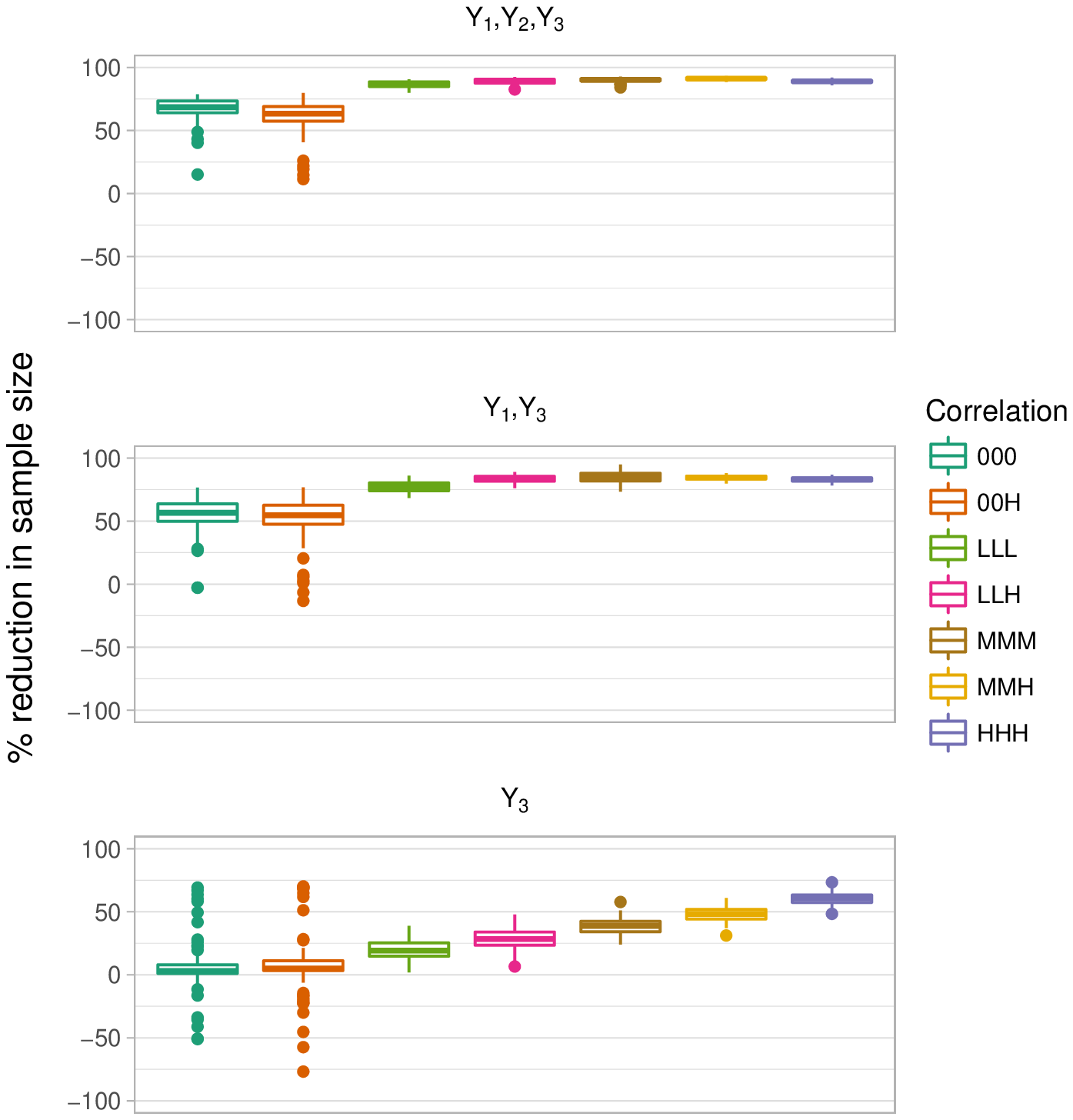}\caption[Boxplots of the estimated reduction in required sample size by using the latent variable methods]{Boxplots of the estimated reduction in required sample size from employing the latent variable method instead of the standard binary method for a range of correlations between one continuous, one ordinal and one binary measure when all outcomes have equal treatment effect. Response is driven by all components in the top panel, the continuous and binary in the middle panel and only the binary in the bottom panel}\label{reduction}
\end{figure}

\noindent Figure \ref{reduction} shows boxplots of the estimated reduction in required sample size when employing the latent variable rather than the standard binary method for a range of correlations between the endpoints. This is shown for the scenarios where response is driven by $(Y_{1},Y_{2},Y_{3})$, $(Y_{1},Y_{3})$ and $(Y_{3})$. In the case where all three components drive response and the correlation between the endpoints is zero, the latent variable method can reduce the sample size by 18-77\%. However when there is correlation between the endpoints, the sample size is reduced by 80-90\%. A similar pattern occurs when response is driven by the continuous and binary components, however there is a small drop in efficiency gains. For example, when the correlation between endpoints is low, the reduction in required sample size drops from 85\% to approximately 77\%, indicating that the ordinal component with 5 levels contributes to the increased precision. When the binary component is the only driver of response and there is no correlation between the endpoints, the median sample size required is the same for both methods. However, when the binary component is the only driver of response and there is correlation present between the endpoints, the latent variable method offers precision gains over the standard binary method. The magnitude of the gain depends on the strength of the correlation between the endpoints, where a higher correlation results in a larger reduction in required sample size. Tables confirming that the empirical power of the method is close to $80\%$ and the sample sizes required for different correlation, response drivers and treatment effect structures for both methods are included in the Appendix.

\subsection{Two Continuous, One Ordinal, One Binary}
\begin{figure}
\centering
\includegraphics[scale=0.9]{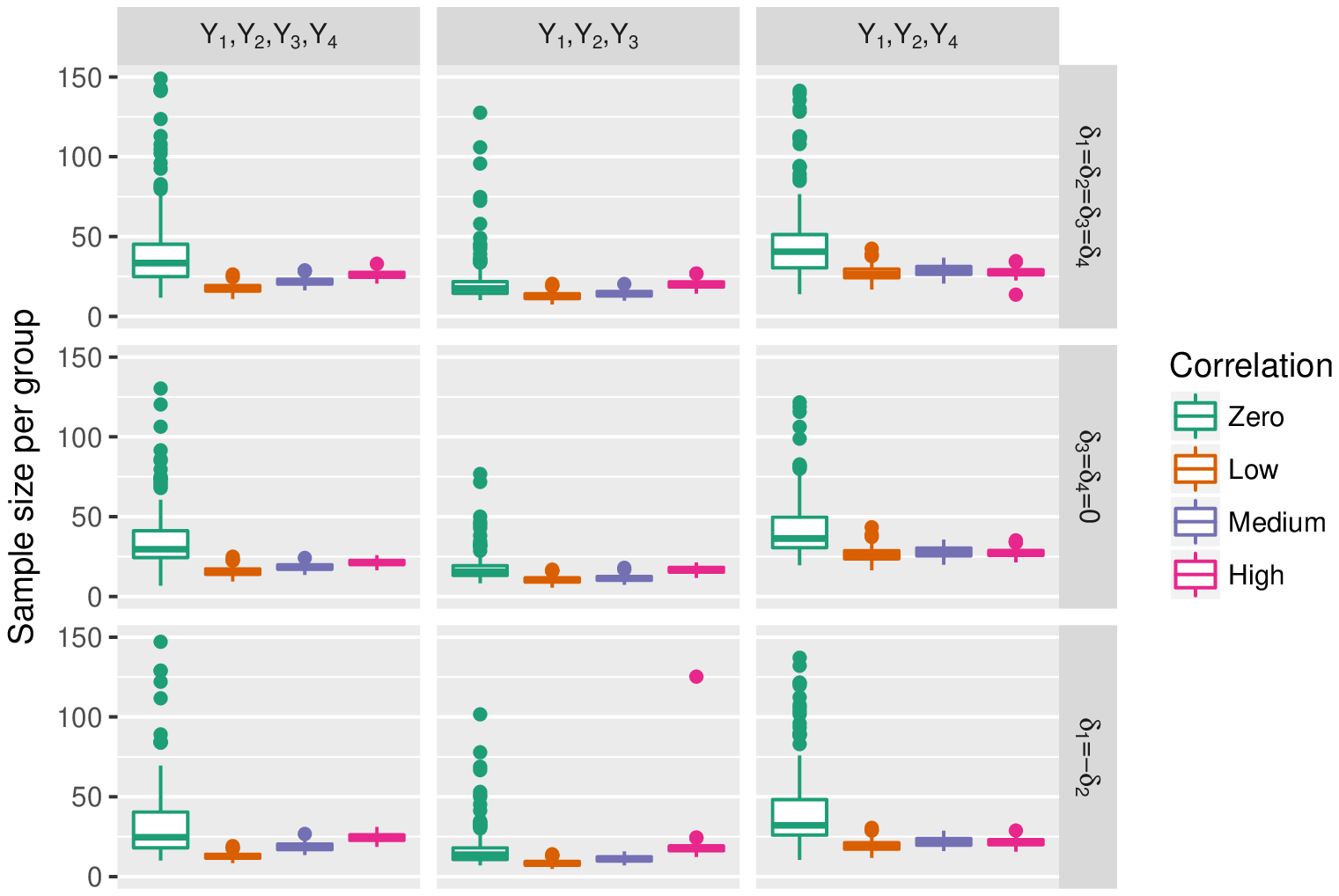}\caption[Boxplots of the estimated sample size per group from the latent variable method for composites containing two continuous, one ordinal and one binary component]{Boxplots of the estimated sample size per group from 1000 simulated datasets using the latent variable method for composites containing two continuous, one ordinal and one binary component and correlation between endpoints is zero, low=0.3, medium=0.5 or high=0.8. These are shown when response is driven by $(Y_{1},Y_{2},Y_{3},Y_{4})$, $(Y_{1},Y_{2},Y_{3})$ or $(Y_{1},Y_{2},Y_{4})$ and the treatment effect structure in the components is $\delta_{1}=\delta_{2}=\delta_{3}=\delta_{4}$, $\delta_{3}=\delta_{4}=0$ or $\delta_{1}=-\delta_{2}$}\label{boxlatsamp}
\end{figure}

As well as considering a composite endpoint made up of one of each type of component, it is interesting to consider the efficiency gains from modelling an additional continuous component. In this instance $Y_{1}$ and $Y_{2}$ are continuous measures, $Y_{3}$ is ordinal and $Y_{4}$ is binary. The results are included in the Appendix, which show that when using the latent variable method, adding a second continuous component which also drives treatment response can reduce the median required sample size by a further 46-58\% for the different correlation structures investigated. The estimated percentage reduction in required sample size from using the latent variable method for these four components is also shown in the Appendix. When the correlation between the components is zero, the median reduction in sample size is 80\%. For any correlation between the outcomes, the median reduction in required sample size is approximately 94\%. Furthermore, the results indicate that most of the efficiency gains are obtained from the continuous measures and only a very small amount of this is from the ordinal variable. \\
Figure \ref{boxlatsamp} shows the sample sizes required across different treatment effect structures, for different components driving response with varying correlation between the components. The sample size required is the same across different treatment effect structures in the components, including when the effects of components are in different directions, as in $\delta_{1}=-\delta_{2}$. The sample sizes required are similar when response is driven by $(Y_{1},Y_{2},Y_{3},Y_{4})$ and $(Y_{1},Y_{2},Y_{3})$. In this setting the sample size is largest for zero correlation and reduces when the components are correlated. 

\section{Discussion}
\label{s:discuss}
The work in this paper demonstrated the various ways in which a latent variable framework may be employed for mixed continuous, ordinal and binary outcomes. We illustrated sample size determination in the case of mixed continuous, ordinal and binary co-primary outcomes. We extended this to allow for sample size determination in the case of mixed multiple primary endpoints and proposed a technique to estimate the sample size when using a latent variable model for the underlying structure of a mixed composite endpoint. Sample size determination in all three cases requires the assumptions of the latent variable framework where discrete variables are treated as continuous variables subject to estimable thresholds. This formulation provides maximum likelihood estimates of latent means and correlation parameters which can be used to characterise the joint distribution. For co-primary and multiple primary endpoints the resulting hypothesis is based on an intersection or union of the hypotheses for the individual outcomes and so is multivariate in nature. However, for composite responder endpoints the hypothesis of interest is stated in relation to the overall responder endpoint and so is univariate. Sample size estimation in this case can make use of the standard power and sample size functions but requires the distribution of the test statistic under the alternative hypothesis which we approximate using a Taylor series expansion. \\
We applied the methods to a numerical example based on a phase IIb study. For the correlation structure observed in the MUSE trial, the sample size required for the co-primary endpoints was similar to but greater than that required for the individual endpoint with the lowest effect size. Alternatively, the sample size required for the multiple primary endpoint changes based on the variance assumed for the outcome with the largest treatment effect, however is lower than that required by the individual endpoint. The sample size required for the composite endpoint was slightly larger than that required for the multiple primary endpoint and similar to that required for the individual endpoint with the largest effect size. Given that in the composite case we are concerned with the overall binary response endpoint, we compared the sample sizes required for the endpoint using the latent variable model with the standard binary method. These findings were in agreement with the power gains found previously \citep{McMenamin2019}. \\
Due to the additional concerns for what drives the sample size requirement in the case of a composite endpoint, we investigated the behaviour of the method under different scenarios in a simulation study. We found that the sample size required depended only on the overall treatment effect in the composite and not the treatment structure in the components. Sample sizes varied depending on the components driving response and the correlation between outcomes. We found that the magnitude of the increase in power offered by the latent variable method is smallest when the components are uncorrelated, however it is still substantial. One important result from this work was quantifying the efficiency gain from adding a second continuous component to the composite, provided both components drive response. We found the median required sample size is reduced by 46-58\% by including the additional continuous component. The results showed that the inclusion of the ordinal component with five levels is only responsible for a very small proportion of the precision gains. Given that the inclusion of the ordinal component substantially increases complexity and computational demand, it may be the case that it is sufficient to combine any ordinal components with the binary outcome. It is likely that the precision gains will be larger for ordinal variables with a larger number of categories however this will greatly increase computation time. Ordinal outcomes with a large number of levels may be included as continuous components. \\
Our results show that the sample sizes required from the standard binary method increase as the correlation between the components increase. These results are unexpected as sample size typically decreases with increasing correlation. For the latent variable method the sample size is largest for zero correlation, as we would expect. However, the sample size required is smaller for low correlation between the components than for medium and high correlation between outcomes. This ambiguity in how the correlation structure affects sample size is problematic for practice. One possible conservative solution is to allow for uncertainty in the correlations and use the maximum required sample size, which will still offer an improvement over the binary method. \\
One practical consideration when calculating the sample size for a trial using the latent variable model is the need to specify a large number of parameters, even in the case of only a few outcomes. Additionally, in order to determine the sample size for composite endpoints we use the delta method to obtain the variance of the risk difference, requiring the covariance matrix of the parameter estimates. This can be obtained by fitting the model to pilot data however this is potentially challenging and restrictive for a number of reasons. Firstly, it requires that a pilot or earlier phase trial must have already taken place in order to apply the method in a certain disease area. This is particularly undesirable in the case of rare diseases which would benefit most from the increased efficiency but where trials are run very infrequently. Furthermore, the pilot data could be fundamentally different to the future trial and observed effects may be imprecise. Therefore, placing too much emphasis on the existing data may lead to problems in the main trial. In theory, it is possible to specify the required covariance parameters without data however this would be difficult in practice. Allowing for uncertainty in the quantities and choosing conservative values should provide an appropriate sample size estimate. An alternative when there is no data available is to apply the method using the sample size required to achieve 80\% power for the binary method and avail of the study having a power much larger than 80\%. \\
It is possible to extend this approach to use adaptive sample size re-estimation, or an internal pilot to allow for reductions in the required sample size in the trial as we collect more information about the treatment effect variability. Future work could also focus on developing the method further to obtain an exact distribution for the test statistic rather that the approximation obtained using the delta method. 

\backmatter

%%%%%% include this section if you wish to acknowledge people,
%%%%%% grant support, etc.

%\section*{Acknowledgements}

%%%%%% include this section only if your manuscript refers to supplementary
%%%%%% materials -- see Instructions for Authors at 
%%%%%% http://www.tibs.org/biometrics

\bibliographystyle{biom} \bibliography{bibliography}

\section*{Supporting Information}

R code and a sample dataset are included as supplementary material.

\clearpage

\section*{Appendix}

\vspace{40mm}

\begin{table}[h]\caption[Empirical power (\%) when employing the latent variable method]{Empirical power (\%) for $n=n_{C}=n_{T}$, $\alpha=0.05$, $\delta=\mu_{T}-\mu_{C}:$ overall risk difference on the composite, $\delta_{1}=\delta_{2}=\delta_{3}$, for a combination of correlations ranging from 0, L=0.3, M=0.5, H=0.8 using the latent variable method when the composite is made up of one continuous, one ordinal and one binary outcome}
\label{emppower}
\begin{tabular}{ccccccccccc}
\toprule
Response & $\delta$ & \multicolumn{3}{c}{n=50} & \multicolumn{3}{c}{n=100} & \multicolumn{3}{c}{n=200}\\
\cmidrule(r){3-5} \cmidrule(r){6-8}\cmidrule(r){9-11} &   & 000 & MMM & HHH  & 000 & MMM & HHH   & 000 & MMM & HHH \\ 
\midrule
$Y_{1},Y_{2},Y_{3}$ & 0.05 & 79.1 & 80.1 & 80.3 & 80.0 & 80.0 & 80.3    &  80.5 &  80.1 & 79.8   \\
 & 0.10 & 80.1& 80.4 & 80.1 & 80.0 & 80.4 & 79.9 &  80.1 & 80.0 & 80.2 \\
& 0.15 & 80.8 & 80.9 & 80.4 &  80.2 & 80.5  & 80.0  & 80.3  & 80.2 & 80.4 \\
$Y_{1},Y_{3}$ & 0.05 & 79.8   & 80.2 & 80.1  & 79.9 & 80.1 & 80.3 & 80.2 & 79.5 & 80.1 \\
& 0.10 & 80.1 & 80.3 & 80.0 & 79.9  & 80.2  & 80.1  & 80.0 & 80.0 &79.8 \\
& 0.15 & 80.2 & 80.4 & 80.3 & 80.1 & 80.2 & 80.7 & 80.3 & 80.1 & 80.0  \\
$Y_{3}$  & 0.05 & 80.1  & 79.7  & 80.2 & 80.1 & 79.2  & 80.4   & 80.0 & 80.1 & 79.9   \\
 & 0.10 & 80.4 & 80.0 & 80.3 & 79.7  & 80.1 & 80.2 & 80.4 & 80.5 & 80.2  \\
 & 0.15  & 80.0 & 80.2  & 80.1 & 80.2  & 80.2 & 80.0  & 80.1 & 80.1 & 80.2 \\
\bottomrule
\end{tabular}\label{emppower}
\end{table}

\clearpage

\begin{figure}[h!]
\centering
\includegraphics[scale=1.1]{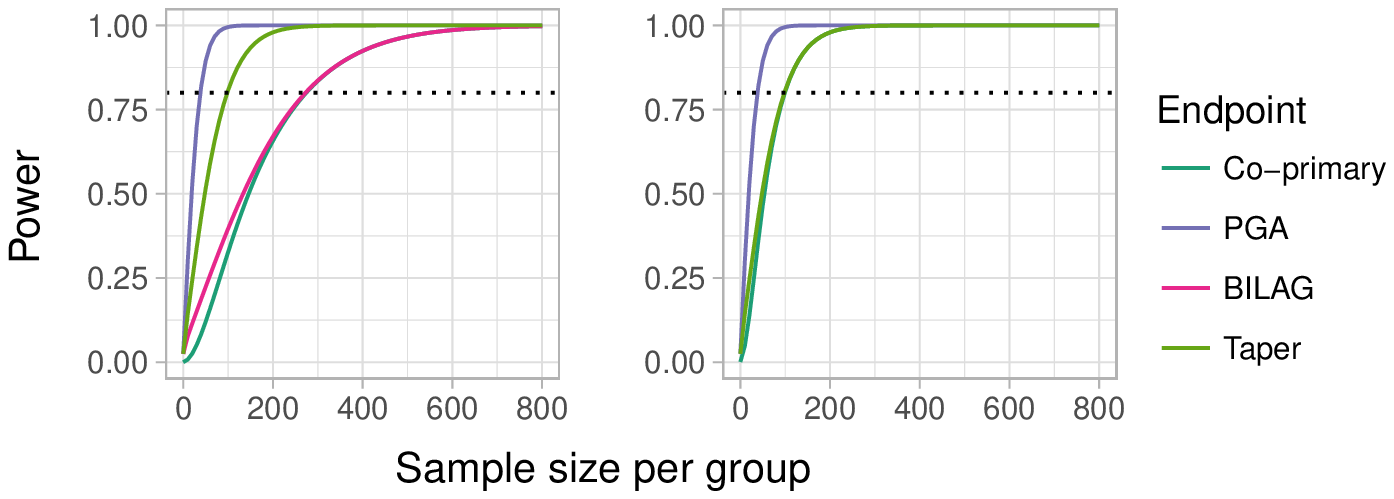}\caption[Power of the co-primary endpoints excluding components]{Overall power $1-\beta$ to detect the treatment effects assumed from the MUSE trial for the systemic lupus erythematosus co-primary endpoints and individual endpoints for different sample sizes per group $n=n_{C}=n_{T}$ for co-primary endpoints PGA, BILAG and Taper (left) and co-primary endpoints PGA and Taper (right)}\label{coprimsample}
\end{figure}

\vspace{20mm}

\begin{figure}[h!]
\centering
\includegraphics[scale=0.9]{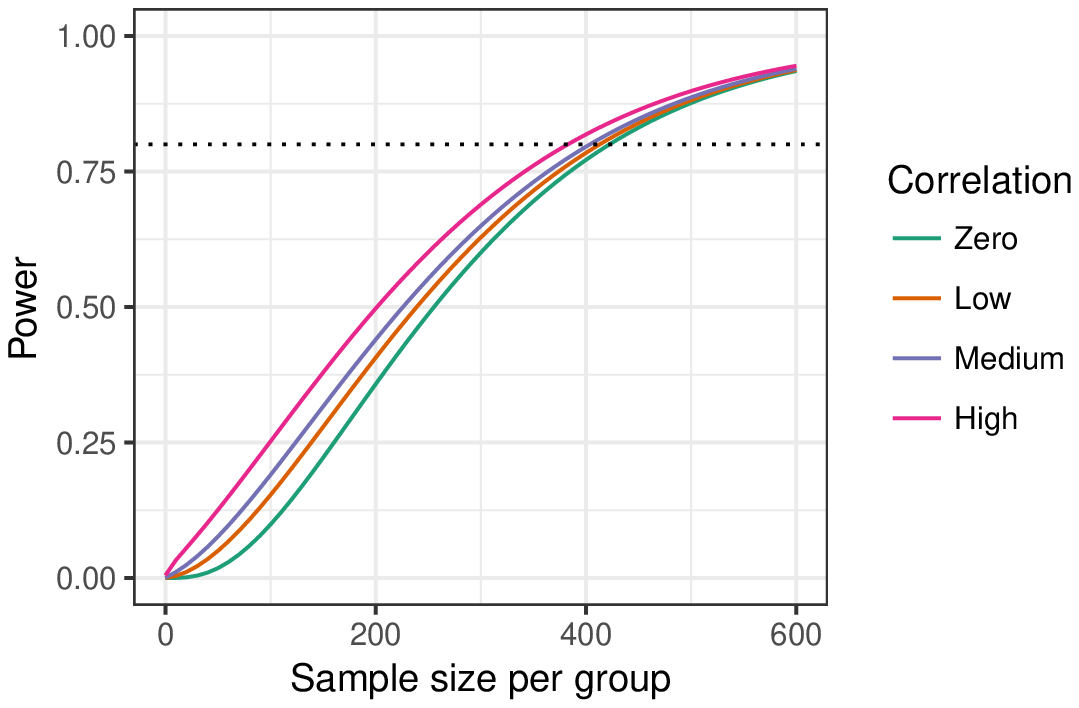}\caption[Power of the co-primary endpoints excluding components for different correlations]{Overall power $1-\beta$ to detect the treatment effects assumed from the MUSE trial for the systemic lupus erythematosus co-primary endpoints for different sample sizes per group $n=n_{C}=n_{T}$ and differing correlations between outcomes, where Low=0.3, Medium=0.5 and High=0.8}\label{coprimcorr}
\end{figure}

\clearpage

\begin{figure}
\centering
\includegraphics[scale=1]{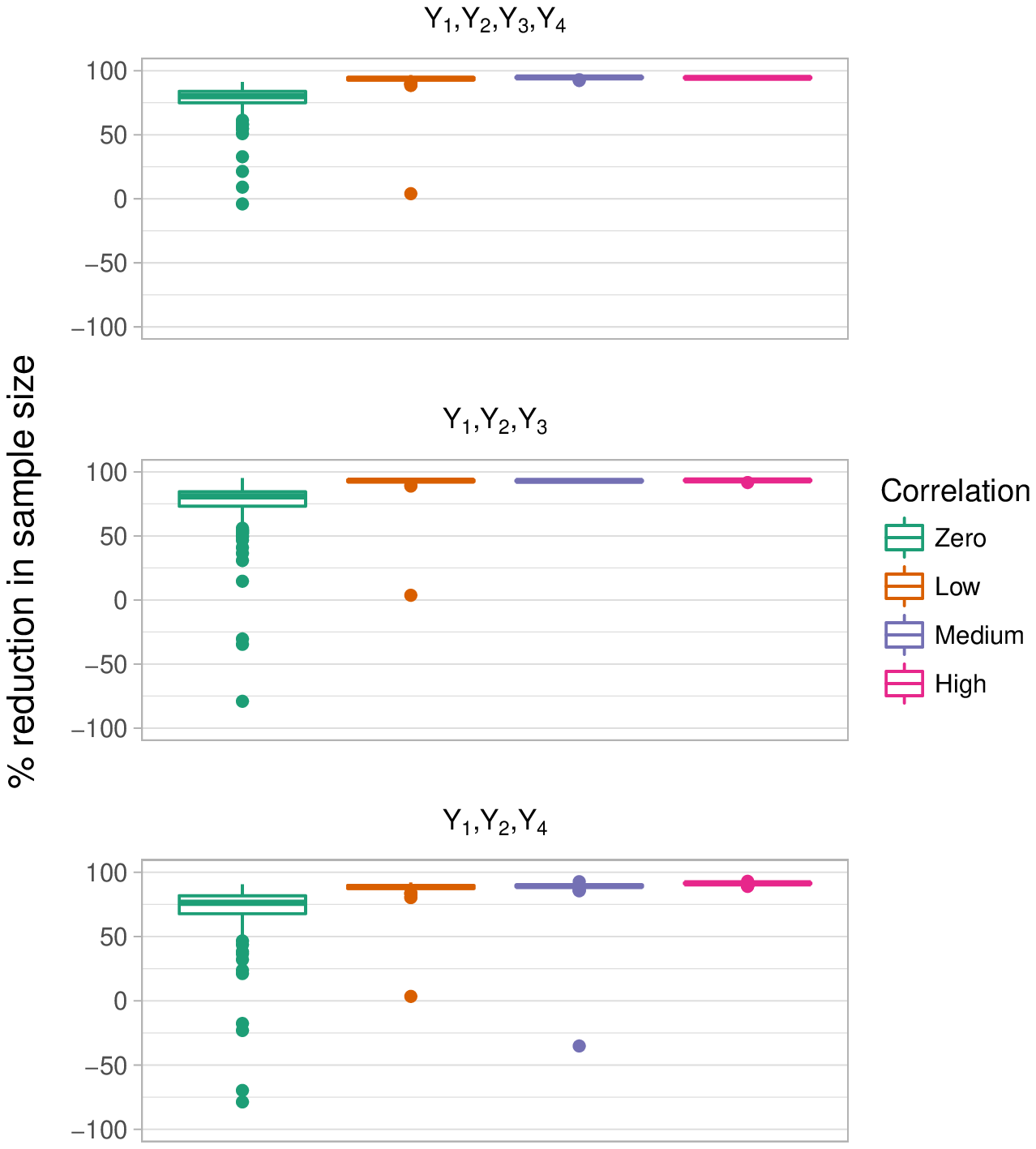}\caption[Boxplots of the estimated reduction in required sample size by using the latent variable method with two continuous components]{Boxplots of the estimated reduction in required sample size in 1000 simulated datasets from employing the latent variable method instead of the standard binary method for correlations of zero, low=0.3, medium=0.5 and high=0.8 between two continuous ($Y_{1},Y_{2}$), one ordinal ($Y_{3}$) and one binary ($Y_{4}$) measure. Response is driven by all components in the top panel, two continuous and ordinal in the middle panel and two continuous and binary in the bottom panel}\label{TCreduction}
\end{figure}

\begin{figure}
\centering
\includegraphics[scale=1]{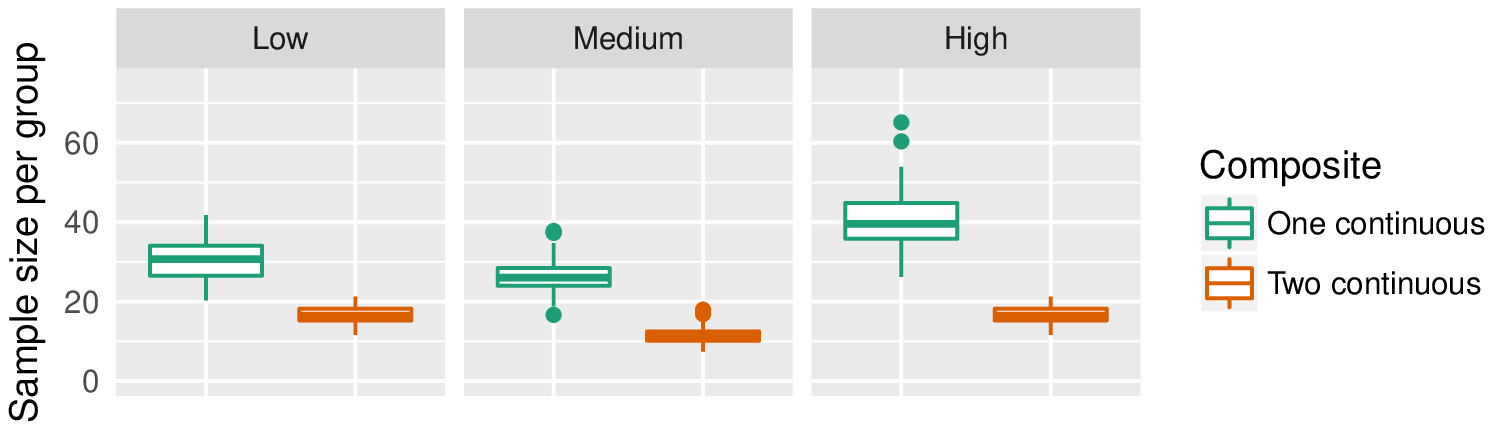}\caption[Boxplots comparing the estimated sample size from the latent variable method for composites containing one and two continuous outcomes]{Boxplots comparing the estimated sample size from 1000 simulated datasets using the latent variable method for composites containing one continuous, one ordinal and one binary component and composites containing two continuous, one ordinal and one binary component, when all outcomes drive response. Correlations between components are low=0.3, medium=0.5 and high=0.8, the risk difference between treatment arms in 0.2 and the treatment effect is the same on all components.}\label{comparisonsamp}
\end{figure}

\begin{table}\caption[Median sample sizes for one continuous, one ordinal and one binary measure using the latent variable method]{Median sample sizes $n=n_{C}=n_{T}$ for overall power $1-\beta \approx 80\%$, $\alpha=0.05$, $k_{m}=k_{o}=K=1$, $\delta=\mu_{T}-\mu_{C}:$ overall risk difference on the composite, $\delta^{\dagger}:$ treatment effect structure in the components, for a combination of correlations 0, L=0.3, M=0.5, H=0.8 using the latent variable model when the composite is made up of one continuous, one ordinal and one binary outcome}
\label{latsampsize}
\begin{tabular}{cccccccccc}
\toprule
Response & $\delta^{\dagger}$ & $\delta$ & \multicolumn{7}{c}{Correlation ($\rho_{12},\rho_{13},\rho_{23}$)}\\
\cmidrule(r){4-10} &   &  & 000 & 00H & LLL & LLH & MMM & MMH & HHH \\ 
\midrule
$Y_{1},Y_{2},Y_{3}$ & $\delta_{1}=\delta_{2}=\delta_{3}$ & 0.05 &  206  & 339  &128  & 119  & 112 & 108 & 145 \\
& & 0.10 & 52 & 85 & 32 & 30 & 28 & 27 & 38 \\
& & 0.15 & 23 & 38 & 15 & 14 & 13 & 12 & 17 \\
& $\delta_{3}=0$& 0.05  &  201 & 334 & 123  &  117 &   105  &   112  &  159   \\
& & 0.10 & 51 & 84 &  31 & 30 & 27  & 28 & 40 \\
& & 0.15 & 23 & 38 & 14 & 13 & 12 & 13 & 18 \\
& $\delta_{2}=\delta_{3}=0$ & 0.05  &  197 & 329 & 119 & 114 & 101 & 106 &  121\\
& & 0.10 & 49 & 81 & 29 & 29 & 27 & 21 & 30 \\
& & 0.15 & 21 & 36& 14 & 13 & 12 & 12 & 15 \\
& & & & & & & & & \\
$Y_{1},Y_{3}$ & $\delta_{1}=\delta_{2}=\delta_{3}$ & 0.05 & 474  & 494 & 289  & 200  & 267 & 205   & 240   \\
& & 0.10 & 119  & 124 & 73 & 50 & 67  & 52 & 60 \\
& & 0.15 & 53 & 55 & 33  & 23 & 30 & 23 & 27 \\
& $\delta_{3}=0$& 0.05  & 468  & 499  &  286  & 195   & 264   &  203   & 248  \\
& & 0.10 & 117 & 125 & 72  & 49 & 66 & 51 & 62  \\
& & 0.15 & 52 & 56 & 32 & 22 & 30 & 23 & 28 \\
& $\delta_{2}=\delta_{3}=0$ & 0.05  &  470 & 501 & 287 & 196 & 264 & 204 & 249  \\
& & 0.10 & 119 & 123 & 73 & 49 & 67 & 51 &  63\\
& & 0.15 & 53 & 55& 32 & 23 & 31 & 24 & 28\\
& & & & & & & & & \\
$Y_{3}$ & $\delta_{1}=\delta_{2}=\delta_{3}$ & 0.05 & 1493  & 1472  &  1250   & 1113   &  948 & 793 & 609   \\
& & 0.10 & 374 & 368 & 313 & 279  & 237  & 199  & 153 \\
& & 0.15 & 166 & 164 & 139 & 124 & 106 & 89 & 68\\
& $\delta_{3}=0$& 0.05  & 1502  & 1468   & 1256  & 1113 & 960  & 806  & 622  \\
& & 0.10 & 376 & 367 & 315  & 279 & 240  & 202 & 156 \\
& & 0.15 & 176 & 164 & 140 & 124 & 107 & 90 & 70 \\
& $\delta_{2}=\delta_{3}=0$ & 0.05  & 1504 & 1465 & 1259 & 1115 & 963 &807 & 624  \\
& & 0.10 & 376  & 370 & 316 & 280& 241 & 203 & 156 \\
& & 0.15 & 174 & 164 & 139 & 126 & 106 & 90 & 70 \\
\bottomrule
\end{tabular}
\end{table}

\begin{table}\caption[Median sample sizes for one continuous, one ordinal and one binary measure using the latent variable method]{Median sample sizes $n=n_{C}=n_{T}$ for overall power $1-\beta \approx 80\%$, $\alpha=0.05$, $k_{m}=k_{o}=K=1$, $\delta=\mu_{T}-\mu_{C}:$ overall risk difference on the composite, $\delta^{\dagger}:$ treatment effect structure in the components, for a combination of correlations ranging from 0, L=0.3, M=0.5, H=0.8 using the standard binary method when the composite is made up of one continuous, one ordinal and one binary outcome}
\label{binsampsize}
\begin{tabular}{cccccccccc}
\toprule
Response & $\delta^{\dagger}$ & $\delta$ & \multicolumn{7}{c}{Correlation ($\rho_{12},\rho_{13},\rho_{23}$)}\\
\cmidrule(r){4-10} &   &  & 000 & 00H & LLL & LLH & MMM & MMH & HHH \\ 
\midrule
$Y_{1},Y_{2},Y_{3}$ & $\delta_{1}=\delta_{2}=\delta_{3}$ & 0.05 & 680 & 965 & 980 & 1141 & 1138   &  1214   &   1352   \\
& & 0.10 & 170 & 242  & 245 & 286 & 285 & 304 & 338  \\
& & 0.15 &  76 & 108 & 109 & 127 & 127 & 135 & 151  \\
& $\delta_{3}=0$& 0.05  & 628  & 939  & 928  & 1098  &  1102  & 1183   &  1332  \\
& & 0.10 & 157 & 235  & 232 & 275 & 276  &  296 & 333 \\
& & 0.15 & 70 & 105 & 104 & 122 & 123 & 132 & 148\\
& $\delta_{2}=\delta_{3}=0$ & 0.05  &  609 & 920 & 914 & 1086 & 1097 & 1171 & 1310  \\
& & 0.10 & 147 & 231 &228  & 270 & 271 & 290 & 328 \\
& & 0.15 & 68 &101 & 101 & 119 & 121 & 130 & 146\\
& & & & & & & & & \\
$Y_{1},Y_{3}$ & $\delta_{1}=\delta_{2}=\delta_{3}$ & 0.05 & 1127   &  1136  & 1255 & 1261  & 1334  & 1320   & 1425   \\
& & 0.10 & 282 & 284  & 314 & 316 & 334 &330  & 357 \\
& & 0.15 & 126 & 127 & 140 & 141 & 149 & 147 & 159\\
& $\delta_{3}=0$& 0.05  & 1078   & 1072   &  1216  & 1218 &  1298   & 1296 & 1403     \\
& & 0.10 & 270 & 268 & 304 & 305  & 325 & 324 & 351 \\
& & 0.15 & 120 & 120 & 136 & 136 & 145 & 144 & 156 \\
& $\delta_{2}=\delta_{3}=0$ & 0.05  & 1066 & 1063 & 1202 & 1209 & 1296 & 1297 & 1403  \\
& & 0.10 & 263  & 259 & 300 &  299 & 319 & 324 & 351 \\
& & 0.15 & 121 &119 & 133 & 131 & 143 & 143 & 156\\
& & & & & & & & &\\
$Y_{3}$ & $\delta_{1}=\delta_{2}=\delta_{3}$ & 0.05 & 1547  & 1548 &  1550  & 1548   & 1550   &  1550  & 1549   \\
& & 0.10 & 387 & 387 & 388 & 387  & 388 & 388 & 388  \\
& & 0.15 & 172 & 172 & 173 & 172 & 173  & 173 & 173\\
& $\delta_{3}=0$& 0.05  &  1544   & 1545 & 1549   & 1548  & 1549   & 1548 &  1545   \\
& & 0.10 & 386 &  387& 388  & 387  & 388 & 387 &  387\\
& & 0.15 & 172 & 172 & 173 & 172 & 173 & 172 & 172 \\
& $\delta_{2}=\delta_{3}=0$ & 0.05  &  1544 & 1546 & 1549 & 1546 & 1551 & 1549 & 1545   \\
& & 0.10 & 386 & 389 & 387 &  385 &  388 & 388 & 387 \\
& & 0.15 & 173 & 172& 173 & 173 & 172 & 172 & 172 \\
\bottomrule
\end{tabular}
\end{table}

\begin{table}\caption[Median sample sizes per group for two continuous, one ordinal and one binary measure using the latent variable method]{Median sample sizes per group $n=n_{C}=n_{T}$ for overall power $1-\beta \approx 80\%$, $\alpha=0.05$, $k_{m}=2, k_{o}=K=1$, $\delta=\mu_{T}-\mu_{C}:$ overall risk difference on the composite , $\delta^{\dagger}$: treatment effect structure in the components, for a combination of correlations 0, L=0.3, M=0.5, H=0.8 using the latent variable model when the composite is comprised of two continuous, one ordinal and one binary outcome}
\label{TClatsampsize}
\begin{adjustbox}{width=1\textwidth}
\begin{tabular}{ccccccc}
\toprule
Response & $\delta^{\dagger}$ & $\delta$ & \multicolumn{4}{c}{Correlation ($\rho_{12},\rho_{13},\rho_{14},\rho_{23},\rho_{24},\rho_{34}$)}\\
\cmidrule(r){4-7} &   &  & 000000 & LLLLLL & MMMMMM & HHHHHH\\ \hline
$Y_{1},Y_{2},Y_{3},Y_{4}$ & $\delta_{1}=\delta_{2}=\delta_{3}=\delta_{4}$ & 0.05 & 70 & 51 & 41 &  81 \\
& & 0.10 & 18 & 13 & 11 & 21 \\
& & 0.15 & 8 & 6 & 5 & 9 \\
& $\delta_{3}=\delta_{4}=0$& 0.05 & 62  & 41 & 47 & 67   \\
& & 0.10 &16 & 11 & 12 & 17\\
& & 0.15 & 7 &5 & 6 & 8 \\
& $\delta_{1}=-\delta_{2}$ & 0.05  & 55 & 34 & 45 & 72 \\
& & 0.10 & 14 & 9 & 12 & 18 \\
& & 0.15 &  7 & 4 & 5 & 8\\
& & & & & & \\
$Y_{1},Y_{2},Y_{3}$ & $\delta_{1}=\delta_{2}=\delta_{3}=\delta_{4}$ & 0.05 & 139 & 71 & 63 & 105 \\
& & 0.10 & 35 & 18 & 16 & 27 \\
& & 0.15 & 16  & 8 & 7 & 12 \\
& $\delta_{3}=\delta_{4}=0$& 0.05 & 120  & 63  & 75  & 85 \\
& & 0.10 & 30 & 16 & 19 & 22\\
& & 0.15 & 13 & 7 & 9 & 10 \\
& $\delta_{1}=-\delta_{2}$ & 0.05  & 105 & 50 & 76 & 99 \\
& & 0.10 & 27 & 13 & 19 & 25 \\
& & 0.15 & 12 & 6 & 9 & 11 \\
& & & & & & \\
$Y_{1},Y_{2},Y_{4}$ &  $\delta_{1}=\delta_{2}=\delta_{3}=\delta_{4}$ & 0.05 & 166   & 106 &  105 & 112 \\
& & 0.10 & 42 &27  & 27 & 28 \\
& & 0.15 & 19 &12 & 12 & 13 \\
& $\delta_{3}=\delta_{4}=0$ & 0.05 & 147   & 105   &113  & 111   \\
& & 0.10 & 37 & 27 & 29 & 28 \\
& & 0.15 & 17 & 12 & 13 & 13\\
& $\delta_{1}=-\delta_{2}$ & 0.05  & 132 & 78 & 88 & 86 \\
& & 0.10 & 33 & 20 & 22 &  22 \\
& & 0.15 &  15 & 9 & 10 & 10 \\
\hline
\end{tabular}
\end{adjustbox}
\end{table}

\begin{table}\caption[Median sample sizes per group for two continuous, one ordinal and one binary measure using the standard binary method]{Median sample sizes per group $n=n_{C}=n_{T}$ for overall power $1-\beta \approx 80\%$, $\alpha=0.05$, $k_{m}=2, k_{o}=K=1$, $\delta=\mu_{T}-\mu_{C}:$ overall risk difference on the composite, $\delta^{\dagger}$: treatment effect structure in the components, for a combination of correlations 0, L=0.3, M=0.5, H=0.8 using the standard binary method when the composite is comprised of two continuous, one ordinal and one binary outcome}
\label{TCbinsampsize}
\begin{adjustbox}{width=1\textwidth}
\begin{tabular}{ccccccc}
\toprule
Response & $\delta^{\dagger}$ & $\delta$ & \multicolumn{4}{c}{Correlation ($\rho_{12},\rho_{13},\rho_{14},\rho_{23},\rho_{24},\rho_{34}$)}\\
\cmidrule(r){4-7} &   &  & 000000 & LLLLLL & MMMMMM & HHHHHH\\ \hline
$Y_{1},Y_{2},Y_{3},Y_{4}$ & $\delta_{1}=\delta_{2}=\delta_{3}=\delta_{4}$ & 0.05 &  386  & 739  & 665 & 1240    \\
& & 0.10 & 97 & 185 & 167 & 310 \\
& & 0.15 & 43  & 83 & 74 & 138 \\
& $\delta_{3}=\delta_{4}=0$& 0.05 &  324 & 665  & 867 & 1205   \\
& & 0.10 & 81 & 167 & 217 & 302\\
& & 0.15 & 36 & 74 & 97 & 134 \\
& $\delta_{1}=-\delta_{2}$ & 0.05  & 331 & 666 & 858  &  1169 \\
& & 0.10 & 83 & 167 & 215 & 293 \\
& & 0.15 & 37 & 74 & 96  & 130 \\
& & & & & & \\
$Y_{1},Y_{2},Y_{3}$ & $\delta_{1}=\delta_{2}=\delta_{3}=\delta_{4}$ & 0.05 & 690  & 956  &  912 & 1300  \\
& & 0.10 & 173 & 239 & 228 & 325 \\
& & 0.15 & 77 & 107 & 102 & 145 \\
& $\delta_{3}=\delta_{4}=0$& 0.05 & 650  & 912  &  1053   & 1283    \\
& & 0.10 & 163 & 228 & 264 & 321\\
& & 0.15 & 73 & 102 & 117 &  143\\
& $\delta_{1}=-\delta_{2}$ & 0.05  & 605 & 866  & 1017  & 1232 \\
& & 0.10 & 152 & 217 & 255 & 308 \\
& & 0.15 &  68 & 97 & 113 & 137\\
& & & & & & \\
$Y_{1},Y_{2},Y_{4}$ &  $\delta_{1}=\delta_{2}=\delta_{3}=\delta_{4}$ & 0.05 & 690  & 962  & 919  & 1298  \\
& & 0.10 & 173 & 241 & 230 & 325 \\
& & 0.15 & 77 & 107 & 103 & 145 \\
& $\delta_{3}=\delta_{4}=0$& 0.05 & 642  & 919 & 1058  & 1281 \\
& & 0.10 & 161 & 230 & 265 & 321 \\
& & 0.15 & 72 &  103 & 118 & 143  \\
& $\delta_{1}=-\delta_{2}$ & 0.05  & 610 & 876 & 1007  &  1225  \\
& & 0.10 & 153 & 219 & 252 & 307 \\
& & 0.15 & 68 & 98 & 112 & 137  \\
\hline
\end{tabular}
\end{adjustbox}
\end{table}

\label{lastpage}

\end{document}